\documentclass[preprint,12pt,3p]{elsarticle}
\usepackage[utf8]{inputenc} 
\usepackage[T1]{fontenc}    
\usepackage{hyperref}       
\usepackage{url}            
\usepackage{booktabs}       
\usepackage{amsfonts}       
\usepackage{nicefrac}       
\usepackage{microtype}      
\usepackage{lipsum}
\usepackage{amsmath,amssymb,graphicx}
\usepackage{float}
\usepackage{siunitx}
\usepackage{array}
\usepackage{lmodern}
\usepackage{amssymb}
\usepackage{xcolor}
\usepackage{multicol}
\usepackage{multirow}
\usepackage{orcidlink}
\usepackage{amsmath,amssymb,exscale}
\usepackage{amsmath}
\usepackage{blindtext}
\usepackage{hyperref}
\journal{Elsevier}
\begin{document}
\begin{frontmatter}
\title{Applications of residues and poles in solutions of some improper integrals evaluation}
\author[IFPI]{José M. De Sousa\corref{author}\,\orcidlink{0000-0002-3941-2382}}
\cortext[author]{I am corresponding author}
\ead{josemoreiradesousa@ifpi.edu.br}

\affiliation[IFPI]{organization={Instituto Federal de Educa\c c\~ao, Ci\^encia e Tecnologia do Piau\'i -- IFPI},
            addressline={Primavera}, 
            city={São Raimundo Nonato},
            postcode={64770-000}, 
            state={Piauí},
            country={Brazil}}



\begin{abstract}
In Physics, we are generally interested in real solutions involving natural phenomena, where knowledge of real functions of real variables is sufficient to obtain physically relevant results. However, the complexity of phenomena associated with nature lead us to situations where it is necessary to extend mathematical knowledge to complex values of complex variables for reasons of completeness and convenience. A relevant factor is that the real numbers do not form a sufficient basis for the representation of the roots of algebraic and polynomial equations, where the knowledge of the behavior of a complex function presents us with a more complete view of its main properties in comparison to the real variables. Therefore, the theoretical basis of complex variables aims to extend the differential and integral calculus in the representation of the domain of complex numbers. We present here the importance of the Theory of Complex Variables as a natural extension of the theory of real functions in solving basic mathematical problems involving improper Integrals, where its solution is obtained through the Theorem of Residues choosing a convenient contour for mathematical solution of some important and useful improper integrals for science and technology students. We hope that the detailed resolutions of some basic improper integrals will contribute to the academic training of students who take the discipline of Complex Variables in the areas of Science, Mathematics and Engineering and also for those who intend to pursue a professional, academic career in Teaching and Research.
\end{abstract}

\end{frontmatter}

\section{Introdução}
\label{INT}

Mathematical Equations of the $2^{\circ}$ degree have their origins around $1700$ years BC, where greek Mathematicians Sumerian used clay tablets with ruler and compass. At this time, greek natural philosophers already obtained square roots of negative numbers in some cases. The Hindus ($1114 - 1185$) assigned the name \textit{Baskara} (hence the origin of the name \textit{Baskara's Formula}) to obtain the roots of an equation of $2^{\circ}$ degree, where in some cases, negative roots appeared. Despite the progress in Algebra, the Hindus and mathematicians of the time defined that the problem had no solution \cite{datta1935hi}. 

Rafael Bombelli $(1523-1573)$ was the first mathematician to study the notation of roots of negative numbers, where he introduced the notation $\sqrt{-1}$, called \textit{piú di meno} and proposed mathematically for the first time instead the mathematical notation of $a + \sqrt{-b}$ and $a - \sqrt{-b}$. \cite{jayawardene1973influence}. René Descartes $(1596-1650)$ in his intellectual production \textit{La Geometrie} wrote in his work that ``\textit{For any equation it is possible to obtain as many roots as the degree allows, but, in many cases, there are no number of roots that corresponds to what we imagine}''. It was through this presentation by René Descartes that we used the nomenclature of imaginary for the mathematical representation $\sqrt{-1}$ until the present day in the Theory of Complex Variables \cite{descartes2012geometry}. It was then Leonhard Euler $(1707-1783)$ who first introduced the notation $i=\sqrt{-1}$ \cite{euler1953leonhard} and Carl Friedrich Gauss $(1777-1855)$ was who introduced the complex number denomination \cite{gauss1874carl}. The basic understanding of a complex number occurs when we try to solve an equation of $2^{\circ}$ degree, where its discriminant, universally represented by the greek letter delta, is negative. A priori, we show that this equation has no solution in the set $\mathbb{R}$ of real numbers, but that there is a set where such solutions exist and which is called the set of complex numbers $\mathbb{C}$. 

So the set of complex numbers is the union of real numbers and imaginary numbers. In this way, we justify the motivations for creating the set $\mathbb{C}$ which was to present the existence of a solution for certain equations of the $2^{\circ}$ degree and extends as an initial introductory basis for Complex Analysis. Therefore, natural phenomena such as the Theory of Heat Transmission, Fluid Mechanics, Electricity, Quantum Mechanics and other phenomena in continuous media, it is necessary to analyze functions of Complex Variables \cite{churchill1960complex}.
In theoretical development in classes taught in Physics, Mathematics, Engineering and other related areas, many problems involve the resolution of integrals. Therefore, integrals of real functions often arise that cannot be solved by the usual methods of integration, so these improper integrals can be solved on the set of complex numbers by performing an Integration in the Complex Plane. Thus, the fundamental theoretical basis of integration in the complex plane is Cauchy's integral theorem \cite{cauchy1884oeuvres,stewart1960numerical}, and also another fundamental mathematical property for Mathematical Physics methods is the residues and poles theorem. The fundamental motivation for studying and understanding the theory of Complex Variables it is in the general properties of functions. The Cauchy-Goursat theorem afirm that if a function is analytic at all points inside and on a simple closed contour $C$, then the value of the improper integral of the function around that contour is zero. However, if the function is not analytic at a finite number of points interior to $C$, there is a specific number, called the residual, with which each of these points contributes to the value of the improper integral in complex plane. The develop of this theory of residues and poles is powerfull method in their developed and solutions in certain areas of applied mathematics and naturals science that involving improper integrals. An understandable way to understand the importance of complex numbers is in the uniqueness of physical problems found in nature, looking at the sources of the singularities found in the problems. Therefore, it is possible, starting from the singularities of a complex function, to completely understand a function. Therefore, it is possible, starting from the singularities of a complex function, to completely understand a function. So, we present here in this theoretical work, the resolution process in details of some fundamental improper integrals usually found in problems involving Physics, Mathematics, Engineering and other related areas.

We hope that the results of the resolutions of the improper integrals presented here will contribute to the training of students of Science and their Technologies in understanding the Theorem of Residues and Poles, choosing convenient contours for the resolution of improper Integrals that cannot be solved by basic intragations methods in the set of real numbers \cite{churchill1960complex}.
\section{Fundamental theorems of integration in the complex plane}

In this section, we present the detailed resolution of some improper Integrals in the Complex Plane, presenting the fundamental mathematical properties of the
complex integrals, such as Cauchy-Goursat's Theorem, Green's Theorem in the Plane and in multiply connected regions and integration boundary deformation \cite{churchill1960complex}.

\subsection{Cauchy-Goursat theorem}

If a function $f (z) = u (x, y) + iv (x, y)$ is analytic at all points of a simply connected domain $\zeta$, then for every closed simple contour $C$ inside from $\zeta$:
\begin{eqnarray}
\oint_{C} f(z)dz = 0
\label{Eq:01}
\end{eqnarray}
where its mathematical proof is given by:
\begin{eqnarray}
    \oint_{C} f(z)dz = \oint_{C} (u + iv) (dx +idy) = \oint_{C} (u + dx - vdy) + i \oint_{C} (vdx + udy) &=&\nonumber \\
 \iint_{D} \left ( \frac{\partial v}{\partial x} + \frac{\partial u}{\partial y}  \right )dxdy + i \iint_{D} \left ( \frac{\partial u}{\partial x} - \frac{\partial u}{\partial y}  \right )dxdy &=& \nonumber \\
 - \iint_{D} \left ( \frac{\partial v}{\partial x} + \frac{\partial v}{\partial x}  \right )dxdy + i \iint_{D} \left ( \frac{\partial u}{\partial x} - \frac{\partial u}{\partial x}  \right )dxdy = 0
 \label{Eq:02}
\end{eqnarray}
where in the domain $\zeta$ being multiply connected, we apply the Cauchy-Goursat theorem \cite{churchill1960complex}.

\subsubsection{In multiply connected regions}

\begin{figure}[htb!]
    \centering
    \includegraphics[scale=0.45]{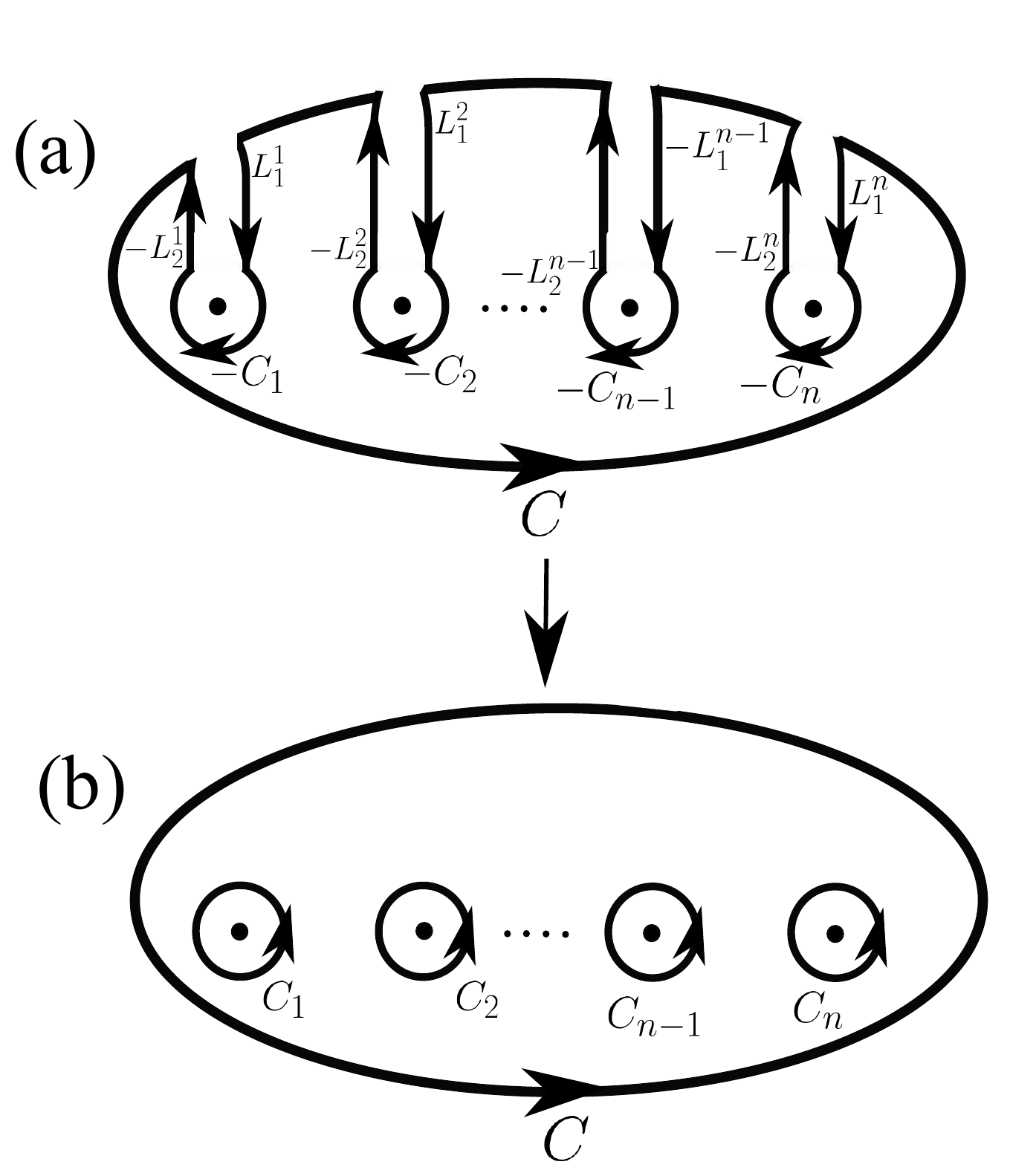}
    \caption{Contour $A$ that transforms a multiply connected region into a simply connected region.}
    \label{Fig:01}
\end{figure}

When $z_{0}$ is an isolated singular point of a function $f$, there exists a positive number in the set of real numbers $N$ such that $f$ is analytic at every point $z$ for which $0 < | z - z_{0}| < N$ . Therefore, $f(z)$ is represented by a Laurent series \cite{monforte2013formal}:
\begin{eqnarray}
f(z) = \sum_{n=0}^{\infty} \alpha_{n} (z - z_{0})^{n} + \frac{\beta_{1}}{(z - z_{0})} + + \frac{\beta_{2}}{(z - z_{0})^{2}} + \frac{\beta_{3}}{(z - z_{0})^{3}} + ... + + \frac{\beta_{n}}{(z - z_{0})^{n}} + .... ~~~ \nonumber \\
(0 < |z -z_{0}| < N),
\label{Eq:03}
\end{eqnarray}
where the coefficients $\alpha_{n}$ and $\beta_{n}$ represent integrals. So, 
\begin{eqnarray}
\beta_{n} = \frac{1}{2 \pi i} \oint_{C} \frac{f(z)}{(z - z_{0})^{(-n +1)}} ~~~ (n = 1, 2, 3, ...., n),
\label{Eq:04}
\end{eqnarray}
where $C$ is some specific closed contour positively oriented around $z_{0}$ and located in $0 < |z - z_{0}| < N$, see Figure \ref{Fig:01}. When $n = 1$, this expression for $\beta_{n}$ can be represented by:
\begin{eqnarray}
 \oint_{C} f(z) dz = 2 \pi i \beta
 \label{Eq:05}
\end{eqnarray}

Therefore, the complex number $\beta$ in Equation \ref{Eq:05} is mathematically called the residue of the function $f(z)$ on an isolated singular point $z_{0}$, so in the mathematical literature the representation is usually used $Res_{z = z_{0}} f(z)$. Consider $C$ as a simple closed contour ($C_{i}, ~(i = 1, 2, 3, ... , n)$) with a finite set of closed contours inside $C$, where there are two regions $Ri$ and $R'_{i}$, interior to $C_{i}$ and $C'_{i}$. Consider $R$ as the region of the complex plane that includes all points inside and along contour $C$, with the exception of points inside each contour $C_{i}$ . Therefore, let $A$ be the complete contour that delimits $R$ consistent with $C$ and the contours $C_{i}$, where all these contours following the sense that keep the points of $R$ to the right of $A$ . So, if the function $f(z)$ (see and Equation \ref{Eq:01}) is analytic in $R$, this presented configuration can be better understood by viewing Figure \ref{Fig:01}.



In Figure \ref{Fig:01} the contour is represented by the letter $C$, the contours $C_{1} ... C_{n}$ and the line segments $L_{1}^{1}, L_ {2}^{1}$ and $L_{2}^{1}, L_{2}^{2}, ... $ and $L_{1}^{n}$ and $L_{2}^ {n}$, whose distance is infinitely small where the path integrals of the complex function $f(z)$ for each pair of segments cancel each other out, where the remaining contour is exactly the contour $A$:
\begin{eqnarray}
\int_{L_{1}^{1}} f(z) dz = - \int_{L_{2}^{1}} f(z) dz, \int_{L_{1}^{2}} f(z) dz = - \int_{L_{2}^{2}} f(z) dz, ... , \int_{L_{1}^{n}} f(z) dz = - \int_{L_{2}^{n}} f(z) dz
\label{Eq:006}
\end{eqnarray}
When $z_{0}$ is an isolated singular point of a function $f$, there exists a positive number in the set of real numbers $N$ such that $f$ is analytic at every point $z$ for which $0 < |z - z_{0}| < N$ . Therefore, $f(z)$ is represented by a Laurent series \cite{monforte2013formal}. Other theorems such as Integration Contour Deformation, Path Independence, Analyticity and Morera Theorem can be researched by students at references, with the aim of complementing their studies in a more general and complete way \cite{churchill1960complex,avila2008variaveis}.
\section{Calculating  improper integrals in the complex plane}

Here we present to Science and Technology students the resolution of some important improper integrals for understanding integrals in the complex plane. We hope that the problems presented below will be useful and motivational for solving several improper integrals that are not solvable within the formalism of the set of real numbers. We present below the calculation in detail of some important improper integrals applying the Residues and Poles Theorem in the Complex Plane.


\subsection*{Applying the residues theorem, choosing a convenient contour, we present below the resolution in detail of some important improper integrals.}

Consider the first improper integral (a), Eq. \ref{Eq:07}:
\begin{eqnarray}
(a) ~~~\int_{0}^{\infty } \frac{x^{2}+1}{x^{4}+1}dx = \frac{\pi}{\sqrt{2}}
\label{Eq:07}
\end{eqnarray}
\textbf{Resolution:}
consider the following integral:
\begin{eqnarray}
\oint_{C} \frac{z^{2}+1}{z^{4}+1}dz
\label{Eq:08}
\end{eqnarray}
\begin{eqnarray}
 Roots: z^{4}+1 = 0 ~ \therefore ~ z^{4} = e^{i(\pi + k2\pi)} ~ \therefore ~ (z^{4})^{\frac{1}{4}} = \left [ e^{i(\pi + k2\pi)} \right ]^{\frac{1}{4}} ~ \therefore ~ z = e^{i\frac{\pi}{4} + k\frac{\pi}{2}} ~ \therefore ~  \nonumber\\
 z = e^{i(\frac{\pi}{4} + k\frac{\pi}{2})} ~ \therefore ~ \boxed{z = e^{i\frac{\pi}{4} + ik\frac{\pi}{2}}}   \nonumber \\
\label{Eq:09}
\end{eqnarray}
So, the roots of the denominator are:
\begin{eqnarray}
e^{i\frac{\pi}{4}};~ e^{i\frac{3\pi}{4}} ;~ e^{i\frac{5\pi}{4}} ; ~ e~ e^{i\frac{7\pi}{4}} .
\label{Eq:010}
\end{eqnarray} 
So we have that the contour is:
\begin{figure}[htb!]
    \centering
    \includegraphics[scale=0.35]{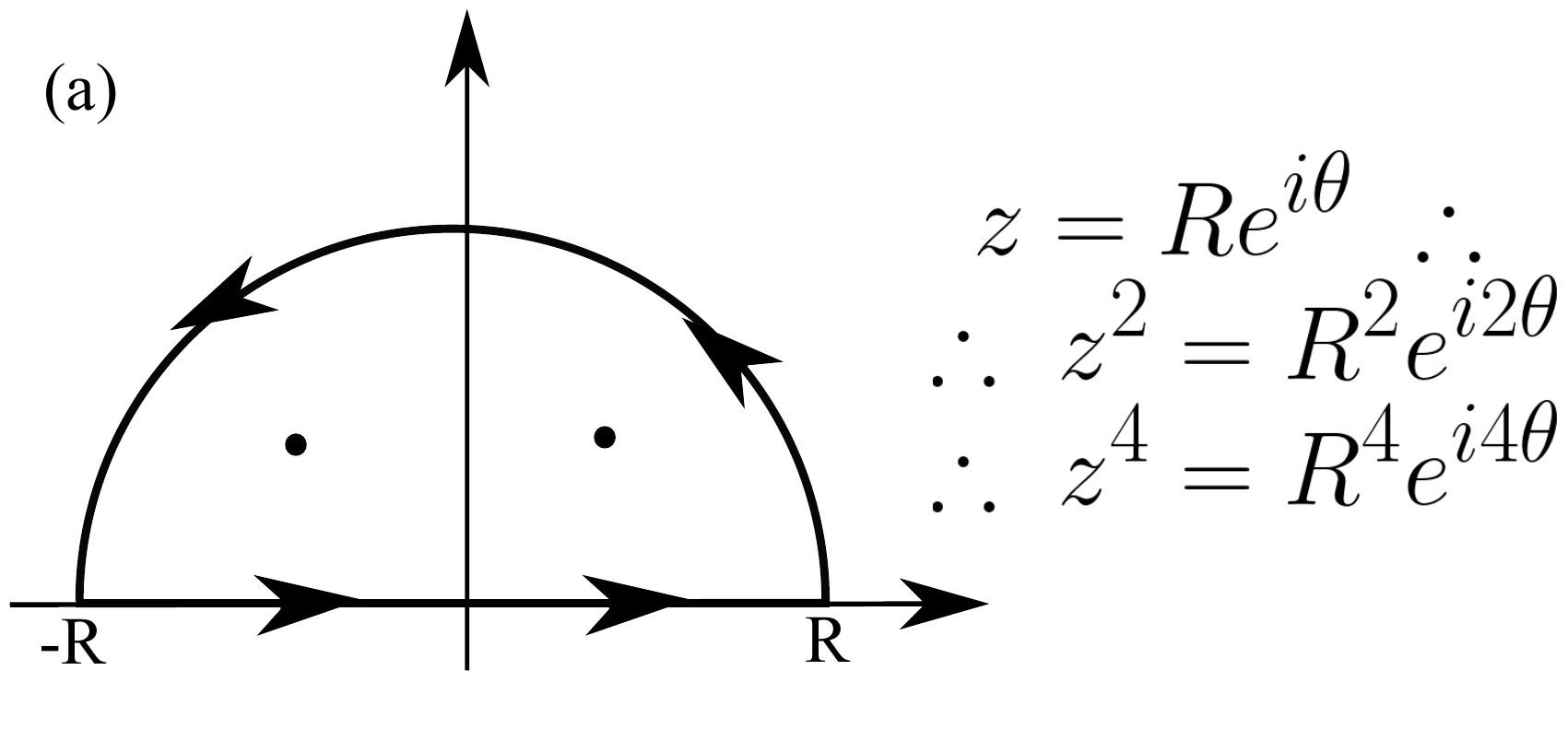}
    \caption{Choosing the convenient contour to solve the Integral \ref{Eq:08}}
    \label{Fig:02}
\end{figure}
so,
\begin{eqnarray}
z = R e^{i\theta}  ~ \therefore ~ \frac{dz}{d\theta} = iRe^{i\theta} ~ \therefore ~ \boxed{dz = iRe^{i\theta}d\theta}   \nonumber \\
\label{Eq:011}
\end{eqnarray}
\begin{eqnarray}
&&\int_{-R}^{R } \frac{x^{2}+1}{x^{4}+1}dx + \int_{0}^{\pi } \frac{z^{2}+1}{z^{4}+1}dz = 2\pi i \sum Res f(z) ~ \therefore ~ \nonumber \\ 
&& ~ \therefore ~ \int_{-R}^{R } \frac{x^{2}+1}{x^{4}+1}dx + \int_{0}^{\pi } \frac{(R^{2}e^{i2\theta}+ 1)iRe^{i\theta}}{(R^{4}e^{14\theta}+1)}d\theta = 2\pi i \sum Res f(z) ~ \therefore ~ \nonumber \\
~ \therefore ~ polos ~~ e^{i\frac{\pi}{4}} ~~and~~ e^{i\frac{3\pi}{4}}
\label{Eq:012}
\end{eqnarray}
Now let's determine the residuals:
\begin{eqnarray}
Res_{z=e^{i\frac{\pi}{4}}} = lim_{z\rightarrow e^{i\frac{\pi}{4}}} \left ( z - e^{i\frac{\pi}{4}} \right ) \left( \frac{z^{2} + 1}{z^{4} + 1} \right )  = \left.\begin{matrix}
\frac{z^{2} + 1}{4 z^{3}} & 
\end{matrix}\right|_{z=e^{i\frac{\pi}{4}}} ~ \therefore ~ \nonumber \\ 
 ~ \therefore ~ Res_{z=e^{i\frac{\pi}{4}}} = \frac{e^{i2\frac{\pi}{4}} + 1}{4 e^{i \frac{3\pi}{4}}} = \frac{e^{1\frac{\pi}{2}}+1}{4 e^{i \frac{3\pi}{4}}} = \frac{(cos(\frac{\pi}{2}) + i sen(\frac{\pi}{2}))+2}{4 \left ( cos (\frac{3\pi}{4}) + i sen (\frac{3\pi}{4}) \right )}  = \frac{(0 + i.1) + 1}{4(-\frac{\sqrt{2}}{2} + i \frac{\sqrt{2}}{2})} \nonumber \\
~ \therefore ~ \boxed{Res_{z=e^{i\frac{\pi}{4}}} = \frac{1 + i}{4\left ( -\frac{\sqrt{2}}{2} + i \frac{\sqrt{2}}{2} \right )}}
\label{Eq:013}
\end{eqnarray}
considering, $e^{i \frac{3\pi}{4}}$, we have to:
\begin{eqnarray}
Res_{z=e^{i \frac{3\pi}{4}}} = lim_{z\rightarrow e^{i\frac{3\pi}{4}}}\left( z - e^{-i \frac{3\pi}{4}} \right)\left( \frac{z^{2} + 1}{z^{4} + 1} \right) = \frac{z^{2} + 1}{4z^{3}}|_{z=e^{i \frac{3\pi}{4}}}   \nonumber \\
~ \therefore ~ Res_{z=e^{i \frac{3\pi}{4}}} =\frac{z^{2} + 1}{4z^{3}}|_{z=e^{i \frac{3\pi}{4}}} 
= \frac{e^{i\frac{3\pi}{2} + 1}}{4e^{i\frac{9\pi}{4}}} = \frac{\left( cos(\frac{3\pi}{4}) + i sen(\frac{3\pi}{2}) + 1) \right)}{4 cos(\frac{9\pi}{4}) + 4i sen(\frac{9\pi}{4}) } = \boxed{\frac{-i + 1}{4\left( \frac{\sqrt{2}}{2} + i\frac{\sqrt{2}}{2}  \right)}}
\label{Eq:014}
\end{eqnarray}
Therefore, the residual is given by:
\begin{eqnarray}
\sum Res f(z) = Res_{z=e^{i\frac{\pi}{4}}} + Res_{z=e^{i\frac{3\pi}{4}}} = \frac{1 + i}{4\left( -\frac{\sqrt{2}}{2} + i \frac{\sqrt{2}}{2} +  \right)} + \frac{1 - i}{4\left( \frac{\sqrt{2}}{2} + i\frac{\sqrt{2}}{2} \right)} = \nonumber \\
=  \frac{4\left( \frac{\sqrt{2}}{2} + i \frac{\sqrt{2}}{2} \right)(1 + i) + 4\left( \frac{-\sqrt{2}}{2} + i \frac{\sqrt{2}}{2}  \right)(1 - i)}{8\left( -\frac{\sqrt{2}}{2} + i \frac{\sqrt{2}}{2}\right) \left(\frac{\sqrt{2}}{2} + i \frac{\sqrt{2}}{2}\right) } =  \frac{\left( \frac{\sqrt{2}}{2} + i \frac{\sqrt{2}}{2} \right)(1 + i) + \left( \frac{-\sqrt{2}}{2} + i \frac{\sqrt{2}}{2}  \right)(1 - i)}{2\left( -\frac{\sqrt{2}}{2} + i \frac{\sqrt{2}}{2}\right)  \left(\frac{\sqrt{2}}{2} + i \frac{\sqrt{2}}{2}\right)} =  \nonumber \\
= \frac{\frac{\sqrt{2}}{2}i - \frac{\sqrt{2}}{2} + \frac{\sqrt{2}}{2} + \frac{\sqrt{2}}{2}i + \frac{\sqrt{2}}{2} i + \frac{\sqrt{2}}{2} - \frac{\sqrt{2}}{2} + \frac{\sqrt{2}}{2} i}{2\left( -1 + i \right)\left( 1 + i \right)} = \frac{2\sqrt{2} i}{2 (-1 -i +i -1)} = \boxed{-\frac{\sqrt{2}}{2}i}
\label{Eq:015}
\end{eqnarray}
Thus, with obtaining the residue, we return to our integral, therefore:
\begin{eqnarray}
\int_{-R}^{R} \frac{x^{2}+1}{x^{4}+1}dx = \int_{0}^{\infty} \frac{(R^{2}e^{i2\theta}+1) i R e^{i\theta}}{(R^{4} e^{i4\theta} + 1)}d\theta = 2\pi i \sum Res f(z) = 2\pi i \left( - \frac{\sqrt{2}}{2} i \right) = \boxed{\pi\sqrt{2}}
\label{Eq:016}
\end{eqnarray}
Considering the limit of Equation \ref{Eq:016} when $R \mapsto \infty$, we therefore have:
\begin{eqnarray}
lim_{R \mapsto \infty} \left( \int_{-R}^{R} \frac{x^{2}+1}{x^{4}+1}dx \right)  = \int_{-\infty}^{\infty} \frac{x^{2}+1}{x^{4}+1}dx
\label{Eq:017}
\end{eqnarray}
thus,
\begin{eqnarray}
\left| \int_{0}^{\pi} \frac{(R^{2}e^{i2\theta} + 1)i e^{i\theta}d\theta}{R^{4}e^{i4\theta} + 1}  \right| \leqslant \left| \int_{0}^{\pi} \frac{(R^{2}e^{i2\theta} + 1)i e^{i\theta}d\theta}{R^{4}e^{i4\theta} + 1}  \right| \nonumber \\
|R^{4}e^{i4\theta} + 1| \leqslant |R^{4}e^{i4\theta}| - 1 = R^{4} - 1 \nonumber \\
\int_{0}^{\pi} \left| \frac{(R^{2}e^{i2\theta}+1) i e^{i\theta}d\theta}{R^{4}e^{i4\theta}+1} \right| \leqslant \int_{0}^{\pi} \frac{R^{2}d\theta}{R^{4} - 1} 
\label{Eq:018}
\end{eqnarray}
therefore, we have the limit:
\begin{eqnarray}
lim_{R \mapsto \infty}\left| \int_{0}^{\pi} \frac{R^{2}d\theta}{R^{4} - 1} \right| \leqslant lim_{R \mapsto \infty} \int_{0}^{\pi} \frac{R^{2}d\theta}{R^{4} - 1}
\label{Eq:019}
\end{eqnarray}
as:
\begin{eqnarray}
0 \leqslant lim_{R \mapsto \infty}\left( \int_{0}^{\pi} \frac{(R^{2}e^{i2\theta} + 1)i e^{i\theta}d\theta}{R^{4}e^{i4\theta} + 1} \right) = 0
\label{Eq:020}
\end{eqnarray}
We can conclude that:
\begin{eqnarray}
\int_{0}^{\pi}  \frac{(R^{2}e^{i2\theta}+1) i e^{i\theta}d\theta}{R^{4}e^{i4\theta}+1} = 0 \nonumber\\
and ~~ 
\int_{-\infty}^{\infty} \frac{x^{2}+1}{x^{4}+1}dx = \pi\sqrt{2} \nonumber \\
or ~~ \nonumber\\
\int_{0}^{\infty} \frac{x^{2}+1}{x^{4}+1}dx = \frac{\pi}{\sqrt{2}}
\label{Eq:021}
\end{eqnarray}

Let us consider the second integral (b), Equation \ref{Eq:022}:
\begin{eqnarray}
(b) ~~~\int_{-\infty}^{\infty } \frac{x}{(x^{2}+4x+13)^{2}}dx = -\frac{\pi}{27}
\label{Eq:022}
\end{eqnarray}

\textbf{Resolution:}
Let's consider the following integral:
\begin{eqnarray}
\oint_{C} \frac{z}{(z^{2}+4z+13)^{2}}dz
\label{Eq:023}
\end{eqnarray}
In the contour shown in Figure \ref{Fig:02}, we will obtain the following roots:
\begin{eqnarray}
z^{2} + 4z + 13 = 0 ~ \therefore ~ \Delta = 16 - 52 = -36 ~ \therefore ~ \sqrt{\Delta} = 6i \nonumber\\
roots: ~~ \boxed{-2 + 3i} ~~ and ~~ \boxed{-2 -3i}
\label{Eq:024}
\end{eqnarray}
So, we have the integral represented in the complex plane:
\begin{eqnarray}
\int_{-R}^{R } \frac{x}{(x^{2}+4x+13)^{2}}dx + \int_{0}^{\pi} \frac{Re^{1\theta}i Re^{i\theta}}{(R^{2}e^{2}e^{i2\theta} + 4 Re^{i\theta} + 13)^{2}}d\theta = 2\pi i Res f(z)
\label{Eq:025}
\end{eqnarray}
The residue is therefore given by:
\begin{eqnarray}
Res f(z) = Res_{z=-2 + 3i} = \left( \frac{z}{(z + 2 + 3i)^{2}} \right)\left.\begin{matrix}~\end{matrix}\right|_{z= -2 + 3i} =  \nonumber \\
= \frac{(z+2+3i)^{2} - 2z(z + 2 + 3i)}{(z+2+3i)^{4}}\left.\begin{matrix}~\end{matrix}\right|_{z= -2 + 3i} = \frac{-36 - 2(-2 + 3i)(6i)}{(6i)^{4}} = \frac{-36 + (4 -6i)(6i)}{6i^{4}} = \nonumber\\
= \frac{-36 + 24i + 36}{6^{4}} ~ \therefore ~ \nonumber\\
\therefore ~ \int_{-R}^{R } \frac{x}{(x^{2}+4x+13)^{2}}dx + \int_{0}^{\pi} \frac{Re^{1\theta}i Re^{i\theta}}{(R^{2}e^{i2\theta} + 4 Re^{i\theta} + 13)^{2}}d\theta = \frac{2\pi i 4i }{2^{3}3^{3}} = \boxed{-\frac{\pi}{27}}
\label{Eq:026}
\end{eqnarray}
Considering the Limit of the Equation \ref{Eq:026}, when $R \to \infty $, we have that:
\begin{eqnarray}
lim_{R \to \infty} \left( \int_{-R}^{R } \frac{x}{(x^{2}+4x+13)^{2}}dx \right) = \int_{-\infty}^{\infty} \frac{x}{(x^{2}+4x+13)^{2}}dx ~ \therefore ~
\label{Eq:027}
\end{eqnarray}
\begin{eqnarray}
~ \therefore ~ \left|  \int_{0}^{\pi} \frac{Re^{1\theta}i Re^{i\theta}}{(R^{2}e^{i2\theta} + 4 Re^{i\theta} + 13)^{2}}d\theta \right| \le \left| \int_{0}^{\pi} \frac{Re^{1\theta}i Re^{i\theta}}{(R^{2}e^{i2\theta} + 4 Re^{i\theta} + 13)^{2}}d\theta \right|
\label{Eq:028}
\end{eqnarray}
\begin{eqnarray}
~ \therefore ~ |(R^{2}e^{i 2\theta} + 4 R e^{i\theta} + 13)^{2}|=|(R^{2}e^{i 2 \theta} + 4 Re^{i\theta} + 13)|^{2} \le ( |R^{2}e^{i2\theta}| - |4 Re^{i\theta}| - 13|)^{2}
\label{Eq:029}
\end{eqnarray}
\begin{eqnarray}
~ \therefore ~ 0 \le lim_{R \to \infty}  \left| \int_{0}^{\pi} \frac{Re^{i\theta}i Re^{i\theta}}{(R^{2}e^{i2\theta} - 4 Re^{i\theta} - 13)^{2}}d\theta \right| \le lim_{R \to \infty} \left| \int_{0}^{\pi}\frac{R^{2}d\theta}{(R^{2} - 4R - 13)^{2}} \right|
\label{Eq:030}
\end{eqnarray}
Therefore, from Equation \ref{Eq:030}, we conclude that:
\begin{eqnarray}
 lim_{R \to \infty}  \left| \int_{0}^{\pi} \frac{Re^{i\theta}i Re^{i\theta}}{(R^{2}e^{i2\theta} - 4 Re^{i\theta} - 13)^{2}}d\theta \right| = 0  ~ \therefore ~ \nonumber\\ 
~ \therefore ~ \int_{-\infty}^{\infty} \frac{x}{(x^{2}+4x+13)^{2}}dx = -\frac{\pi}{27}
\label{Eq:031}
\end{eqnarray}

Let us consider the third integral (c), Equation \ref{Eq:032}:
\begin{eqnarray}
(c) ~~~\int_{0}^{\infty } \frac{dx}{(x^{2}+ 1)^{n}} = \frac{1, 3, 5, ... , (2n - 3)\pi}{2, 4, 6, ..., (2n - 2)2}, ~~ se ~~ n > 1; ~ = \frac{\pi}{2}, ~~ se ~~ n=1; ~~ (n = 1, 2, 3, ..., n)
\label{Eq:032}
\end{eqnarray}

\textbf{Resolution:}
Let's consider the following integral:
\begin{eqnarray}
\oint_{C} \frac{d z}{(z^{2} + 1)^{n}}
\label{Eq:033}
\end{eqnarray}
In the contour shown in Figure \ref{Fig:02}, we will obtain the poles of order $n$, therefore:\\
\begin{eqnarray}
\int_{-R}^{R} \frac{d x}{(x^{2} + 1)^{n}} + \int_{0}^{\pi} \frac{i R e^{i\theta}d\theta}{(R^{2}e^{2i\theta} + 1)^{n}} = 2 \pi i Res f(z)
\label{Eq:034}
\end{eqnarray}
where the residue is:
\begin{eqnarray}
Res = \frac{1}{2z} |_{z = i} = \frac{1}{2i}, 
\label{Eq:035}
\end{eqnarray}
for $n=1$, we therefore have:
\begin{eqnarray}
lim_{R \to \infty} \left( \int_{-R}^{R} \frac{d x}{(x^{2} + 1)} \right) = \int_{-\infty}^{\infty} \frac{d x}{x^{2} + 1} ~ \therefore ~ \nonumber\\ 
\therefore ~ \left| \int_{0}^{\pi} \frac{i R e^{i\theta}d\theta}{R^{2}e^{2i\theta} + 1}  \right| \le \int_{0}^{\pi} \frac{R d\theta}{R^{2} - 1} ~ \therefore ~ \int_{0}^{\pi} \frac{R d\theta}{R^{2} - 1} = 0 ~ \therefore ~ \nonumber \\
0 \le lim_{R \to \infty} \left| \int_{0}^{\pi} \frac{i R e^{i\theta}d\theta}{R^{2}e^{2i\theta} + 1}  \right| \le lim_{R \to 0} \left( \int_{0}^{\pi} \frac{R d\theta}{R^{2} - 1}\right) ~ \therefore ~ \nonumber \\
~ \therefore ~ lim_{R \to \infty} \left( \int_{0}^{\pi}  \frac{i R e^{i\theta}d\theta}{R^{2}e^{2i\theta} + 1}  \right) = 0 ~ \therefore ~ \int_{-\infty}^{\infty} \frac{d x }{x^{2} + 1} = 2 \pi i \frac{1}{2i} = \pi ~ \therefore ~ \int_{0}^{\infty} \frac{d x }{x^{2} + 1} = \frac{\pi}{2}
\label{Eq:036}
\end{eqnarray}
for $n=2$, we therefore have:
\begin{eqnarray}
\int_{-R}^{R} \frac{d x}{(x^{2} + 1)^{2}} + \int_{0}^{\pi} \frac{i R e^{i\theta}d\theta}{(R^{2}e^{2i\theta} + 1)^{2}} = 2 \pi i Res f(z) ~ \therefore ~ \nonumber \\
~ \therefore ~ Res_{z=i} = \left( \frac{1}{(z+i)^{2}} \right)|_{z=i} = -\frac{2}{(z + i)^{3}}|_{z=i} = -\frac{2}{(2 i)^{3}} = \frac{-2}{-8i} = \frac{1}{4i} ~ \therefore ~ \nonumber\\
0 \le lim_{R \to \infty} \left| \int_{0}^{\pi} \frac{i R e^{i\theta}d\theta}{(R^{2}e^{2i\theta} + 1)^{2}}  \right| \le lim_{R \to \infty}  \left( \int_{0}^{\pi} \frac{ R d\theta}{(R^{2} - 1)^{2}}\right) = 0 ~ \therefore ~ \nonumber\\
~ \therefore ~ lim_{R \to \infty} \left| \int_{0}^{\pi} \frac{i R e^{i\theta}d\theta}{(R^{2}e^{2i\theta} + 1)^{2}}  \right| = 0 ~ \therefore ~ lim_{R \to \infty} \left| \int_{-R}^{R} \frac{dx}{(x^{2} + 1)^{2}}  \right| =  \int_{-\infty}^{\infty} \frac{dx}{(x^{2} + 1)^{2}} 
\label{Eq:037}
\end{eqnarray}
thus:
\begin{eqnarray}
\int_{-\infty}^{\infty} \frac{dx}{(x^{2} + 1)^{2}} = \boxed{\frac{\pi}{2}} ~~ e ~~ \nonumber\\
\int_{0}^{\infty} \frac{dx}{(x^{2} + 1)^{2}} = \boxed{\frac{\pi}{4}}
\label{Eq:038}
\end{eqnarray}
for $n=3$, we therefore have:
\begin{eqnarray}
\int_{-R}^{R} \frac{d x}{(x^{2} + 1)^{3}} + \int_{0}^{\pi} \frac{i R e^{i\theta}d\theta}{(R^{2}e^{2i\theta} + 1)^{3}} = 2 \pi i Res f(z) ~ \therefore ~ \nonumber \\
lim_{R \to \infty} \left| \int_{-R}^{R} \frac{d x}{(x^{2} + 1)^{3}}  \right| = \int_{-\infty}^{\infty} \frac{d x}{(x^{2} + 1)^{3}} ~ \therefore ~ \nonumber \\
~ \therefore ~ \left| \int_{0}^{\pi} \frac{i R e^{i\theta}d\theta}{(R^{2}e^{2i\theta} + 1)^{3}} \right| \le  \int_{0}^{\pi} \left| \frac{i R e^{i\theta}d\theta}{(R^{2}e^{2i\theta} + 1)^{3}} \right| \le \int_{0}^{\pi} \frac{R d\theta}{(R^{2} - 1)^{3}} ~ \therefore ~ lim_{R \to \infty} \left| 
 \frac{R d\theta}{(R^{2} - 1)^{3}} \right|=0 \nonumber \\
 ~ \therefore ~ 0 \le lim_{R \to \infty}  \left| \int_{0}^{\pi} \frac{i R e^{i\theta}d\theta}{(R^{2}e^{2i\theta} + 1)^{3}} \right| \le lim_{R \to \infty} \left| \int_{0}^{\pi} \frac{R d\theta}{(R^{2} - 1)^{3}} \right| ,~~ so: ~~ lim_{R \to \infty} \left| \frac{i R e^{i\theta}d\theta}{(R^{2}e^{2i\theta} + 1)^{3}} \right| = 0
 \label{Eq:039}
\end{eqnarray}
so the residuals are:
\begin{eqnarray}
Res f(z) = \frac{1}{(-3)(-4)}{(z - i )^{5}} = \frac{3.2}{(2i)^{5}} = \frac{3}{16 i} ~ \therefore ~ \nonumber \\
~ \therefore ~ \int_{-\infty}^{\infty} \frac{dx}{(x^{2} + 1)^{3}} = \frac{3\pi}{8} ~ \therefore ~ \boxed{\int_{0}^{\infty} \frac{dx}{(x^{2} + 1)^{3}} = \frac{1.3}{2.4}\frac{\pi}{2}}
\label{Eq:040}
\end{eqnarray}
for $n=4$, we therefore have:
\begin{eqnarray}
\int_{-R}^{R} \frac{d x}{(x^{2} + 1)^{4}} + \int_{0}^{\pi} \frac{i R e^{i\theta}d\theta}{(R^{2}e^{2i\theta} + 1)^{4}} = 2 \pi i Res f(z) ~ \therefore ~ \nonumber \\
 \left| \int_{0}^{\pi}  \frac{i R e^{i\theta}d\theta}{(R^{2}e^{2i\theta} + 1)^{4}}  \right| \le \int_{0}^{\pi} \left| \frac{i R e^{i\theta}d\theta}{(R^{2}e^{2i\theta} + 1)^{4}}  \right| \le \int_{0}^{\pi} \frac{R d\theta}{R^{2} - 1} ~ \therefore ~ lim_{R \to \infty} \int_{0}^{\pi} \left( \frac{R d\theta}{R^{2} - 1} \right) = 0   ~ \therefore ~ \nonumber \\
 ~ \therefore ~ 0 \le lim_{R \to \infty} \left| \int_{0}^{\pi}  \frac{i R e^{i\theta}d\theta}{(R^{2}e^{2i\theta} + 1)^{4}} \right| \le lim_{R \to \infty} \left| \frac{Rd\theta}{R^{2} - 1} \right| ~ \therefore ~  lim_{R \to \infty} \left( \int_{0}^{\pi} \frac{i R e^{i\theta}d\theta}{(R^{2}e^{2i\theta} + 1)^{4}} \right) = 0 \nonumber\\
 lim_{R \to \infty} \left( \int_{0}^{\pi} ~ \therefore ~ \frac{R d\theta}{R^{2} - 1}\right) = 0 ~ \therefore ~ 0 \le  lim_{R \to \infty} \left| \int_{0}^{\pi}  \frac{i R e^{i\theta}d\theta}{(R^{2}e^{2i\theta} + 1)^{4}} \right| \le lim_{R \to \infty} \left| \frac{R d\theta}{R^{2} - 1} \right| ~ \therefore ~ \nonumber \\
 lim_{R \to \infty} \left( \frac{i R e^{i\theta}d\theta}{(R^{2}e^{2i\theta} + 1)^{4}} \right) = 0 ~ \therefore ~ \int_{-R}^{R} \frac{dx}{(x^{2} + 1)^{4}} = \int_{-\infty}^{\infty} \frac{dx}{(x^{2} + 1)^{4}} ~ \therefore ~ \int_{-\infty}^{\infty} \frac{dx}{(x^{2} + 1)^{4}} = 2 \pi i Res f(z) \nonumber\\
 ~ \therefore ~ Res_{z=i} f(z) = \frac{(-4)(-5)(-6)}{(z + i)^{7}}\frac{1}{3!}|_{z=i} = \frac{4.5.6}{2^{8}.3.i} = \frac{1.3.5}{2^{5}.3.i} = \frac{4.5.6}{2^{8}.3.i} = \frac{1.3.5}{2.2^{2}.2.3.2.i} = \frac{1.3.5}{2.4.6} = \frac{1}{2i} \nonumber \\
~ \therefore ~ \int_{-\infty}^{\infty} \frac{dx}{(x^{2}+ 1)^{4}} = \frac{1.3.5}{2.4.6}.\pi ~ \therefore ~ \int_{-\infty}^{\infty} \frac{dx}{(x^{2}+ 1)^{4}} = \frac{1.3.5}{2.4.6}.\frac{\pi}{2} 
\label{Eq:041}
\end{eqnarray}
therefore, for order $n = 1, 2, 3, 4, ... n$, we have:
\begin{eqnarray}
\int_{-\infty}^{\infty} \frac{dx}{(x^{2} + 1)^{n}} = \frac{1.3.5., ..., (2n -3)}{1.3.5., ..., (2n -2)}\frac{\pi}{2}
\label{Eq:042}
\end{eqnarray}

Let's consider the fourth integral (d), Equation \ref{Eq:043}:
\begin{eqnarray}
(d) ~~~\int_{-\infty}^{\infty } \frac{cos(a x)}{(x^{2}+ b^{2})}dx = \frac{\pi e^{-|ab|}}{|b|}
\label{Eq:043}
\end{eqnarray}

\begin{figure}[htb!]
    \centering
    \includegraphics[scale=0.35]{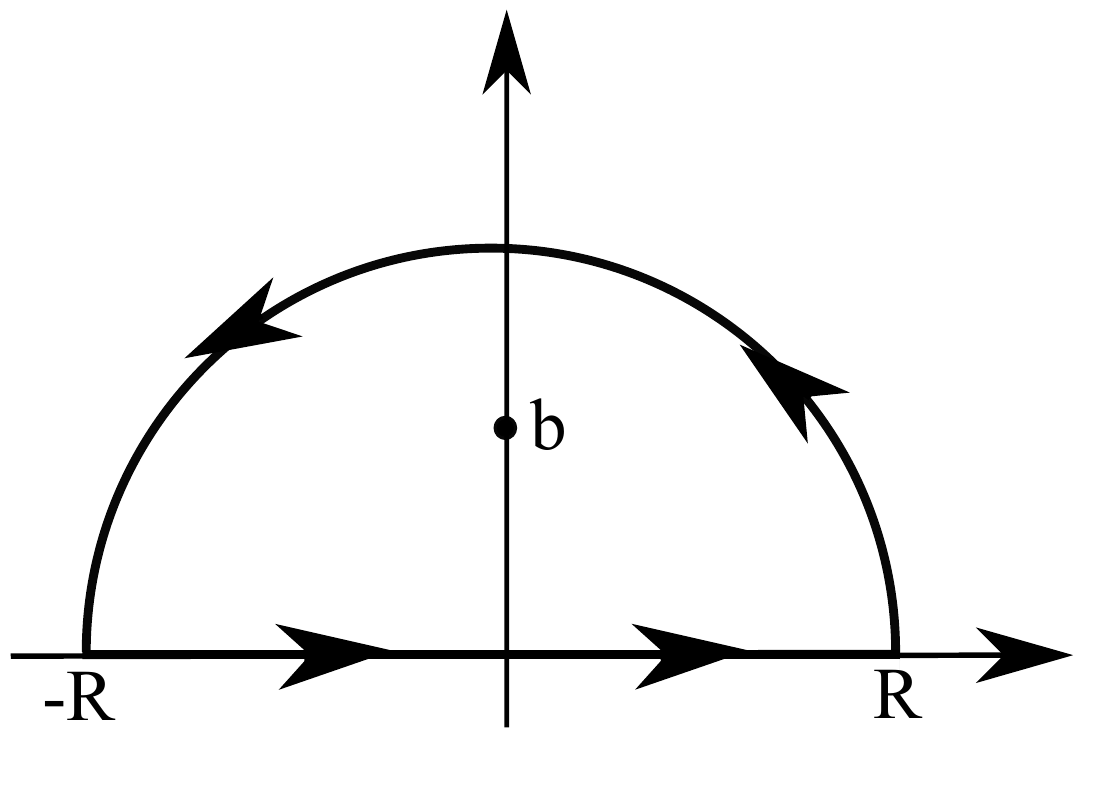}
    \caption{Choice of convenient contour for solving the Integral \ref{Eq:043}}
    \label{Fig:03}
\end{figure}

\textbf{Resolution:}
Let's consider the following integral:
\begin{eqnarray}
\oint_{C} \frac{e^{i a z}}{(z^{2} + b^{2})}dz
\label{Eq:044}
\end{eqnarray}
In the contour shown in Figure \ref{Fig:03}, considering $(a > 0)$, we have a single pole, $b_{i}$ (considering $b > 0$), so we have:\\
\begin{eqnarray}
\int_{-R}^{R} \frac{e^{iax}}{(x^{2} + b^{2})}dx + \int_{0}^{\infty} \frac{e^{iaRe^{i\theta}}iRe^{i\theta}d\theta}{(R^{2}e^{i2\theta} + b^{2})} = 2\pi i Res f(z) \nonumber \\
~ \therefore ~ Res f(z) = lim_{z \to b_{i}} \left[  (z - b_{i}) \frac{e^{iaz}}{(z^{2} + b^{2})}\right] = \frac{e^{iaz}}{2z}|_{z=b_{i}} = \frac{e^{iab_{i}}}{2b_{i}} = \frac{e^{-ab_{i}}}{2b_{i}},
\label{Eq:045}
\end{eqnarray}
as we are considering $a > 0$ and $b > 0$, for consistency of notation we have:
\begin{eqnarray}
Res f(z) = \frac{e^{-|ab|}}{2|b|} ~ \therefore ~ \int_{-R}^{R} \frac{e^{iax}dx}{(x^{2} + b^{2})} + \int_{0}^{\pi} \frac{e^{iaRe^{i\theta}}iRe^{i\theta}d\theta}{(R^{2}e^{i2\theta} + b^{2})} = \frac{\pi e^{-|ab|}}{|b|},
\label{Eq:046}
\end{eqnarray}
we consider the limit of the above expression to be $R \to \infty$, so:
\begin{eqnarray}
lim_{R \to \infty} \left( \int_{-R}^{R} \frac{e^{iax}dx}{(x^{2} + b^{2})} \right) = \int_{-\infty}^{\infty} \frac{e^{iax}dx}{(x^{2} + b^{2})} ~ \therefore ~ |R^{2}e^{i2\theta} + b^{2}| \ge \left| |R^{2}e^{i2\theta}| - |b^{2}|  \right| = |R^{2} - b^{2}| \nonumber \\
~ \therefore ~ \left| e^{iaRe^{i\theta}}  \right| = \left| e^{iaR(cos(\theta) + i sen(\theta))}  \right| = \left| e^{iaRcos(\theta)e^{-sen(\theta)}} \right| = e^{-sen(\theta)} ~ \therefore ~ \nonumber \\
~ \therefore ~ \int_{0}^{\pi} \left| \frac{e^{iaRe^{i\theta}}iRe^{i\theta}d\theta}{(R^{2}e^{i2\theta} + b^{2})}  \right| \le \int_{0}^{\pi} \frac{R e^{-sen(\theta)}d\theta}{|R^{2} - b^{2}|} ~ \therefore ~ lim_{R \to \infty} \left(  \int_{0}^{\pi} \frac{R e^{-sen(\theta)}d\theta}{|R^{2} - b^{2}|} \right) = \nonumber \\
~ \therefore ~ lim_{R \to \infty} \left(  \int_{0}^{\pi} \frac{R e^{-sen(\theta)}d\theta}{R^{2} - b^{2}} \right) ~ \therefore ~ lim_{R \to \infty} \left(  \int_{0}^{\pi} \frac{R e^{-sen(\theta)}d\theta}{|R^{2} - b^{2}|}\right) \le lim_{R \to \infty} \left(  \int_{0}^{\pi} \frac{R e^{-sen(\theta)}d\theta}{R^{2} - b^{2}} \right), 
\label{Eq:047}
\end{eqnarray}
we therefore have:
\begin{eqnarray}
\int_{0}^{\pi} \frac{e^{iaRe^{i\theta}}iRe^{i\theta}d\theta}{R^{2}e^{i2\theta} + b^{2}} = 0 ~~ e ~~ \int_{-\infty}^{\infty} \frac{e^{iax}dx}{(x^{2} + b^{2})} = \frac{\pi e^{-|ab|}}{|b|}, ~~we ~~ get ~~ \boxed{\int_{-\infty}^{\infty} \frac{sen(a x) dx}{(x^{2} + b^{2})} = 0} ~~ and ~~ \nonumber \\
\boxed{  \int_{-\infty}^{\infty} \frac{cos(ax) dx}{(x^{2} + b^{2})} = \frac{\pi e^{-|ab|}}{|b|}}
\label{Eq:048}
\end{eqnarray}

Let's consider the fifth integral (e), Equation \ref{Eq:049}:
\begin{eqnarray}
(e) ~~~\int_{-\infty}^{\infty } \frac{x sen(x)}{(x^{2}+ a^{2})}dx = \frac{\pi}{a}e^{-a}
\label{Eq:049}
\end{eqnarray}

\textbf{Resolution:}
Let's consider the following integral:
\begin{eqnarray}
\oint_{C} \frac{e^{i a z}}{(z^{2} + b^{2})}dz,
\label{Eq:050}
\end{eqnarray}
considering the integral \ref{Eq:048}, already defined here in this work, we have:
\begin{eqnarray}
(e) ~~~\int_{-\infty}^{\infty } \frac{cos(ax)}{(x^{2}+ b^{2})}dx = \frac{\pi e^{-|ab|}}{|b|}, 
\label{Eq:051}
\end{eqnarray}
differentiating with respect to the variable $a$, we get:
\begin{eqnarray}
(e) ~~~ - \int_{-\infty}^{\infty } \frac{xsen(ax)}{(x^{2}+ b^{2})}dx = - \frac{|b|\pi e^{-|ab|}}{|b|}, ~~ for ~~ a=1, ~~ \int_{-\infty}^{\infty } \frac{xsen(ax)}{(x^{2}+ b^{2})}dx =  \pi e^{-b} ~ \therefore ~\nonumber \\
~ \therefore ~ ~~ for ~~ a=b, ~~ \boxed{ \int_{-\infty}^{\infty } \frac{xsen(x)}{(x^{2}+ a^{2})}dx =  \pi e^{a}}
\label{Eq:052}
\end{eqnarray}

Let's consider the sixth integral (f), Equation \ref{Eq:053}:
\begin{eqnarray}
(f) ~~~\int_{0}^{\infty } \frac{x^{a}}{(x^{2}+ 1)^{2}}dx = \frac{\pi a}{sen(a \pi)}, ~~ com: ~~ -1 < a < 1,
\label{Eq:053}
\end{eqnarray}
where to solve the integral (Equation \ref{Eq:054}), consider the following contour (Figure \ref{Fig:04}):
\begin{figure}[htb!]
    \centering
    \includegraphics[scale=0.45]{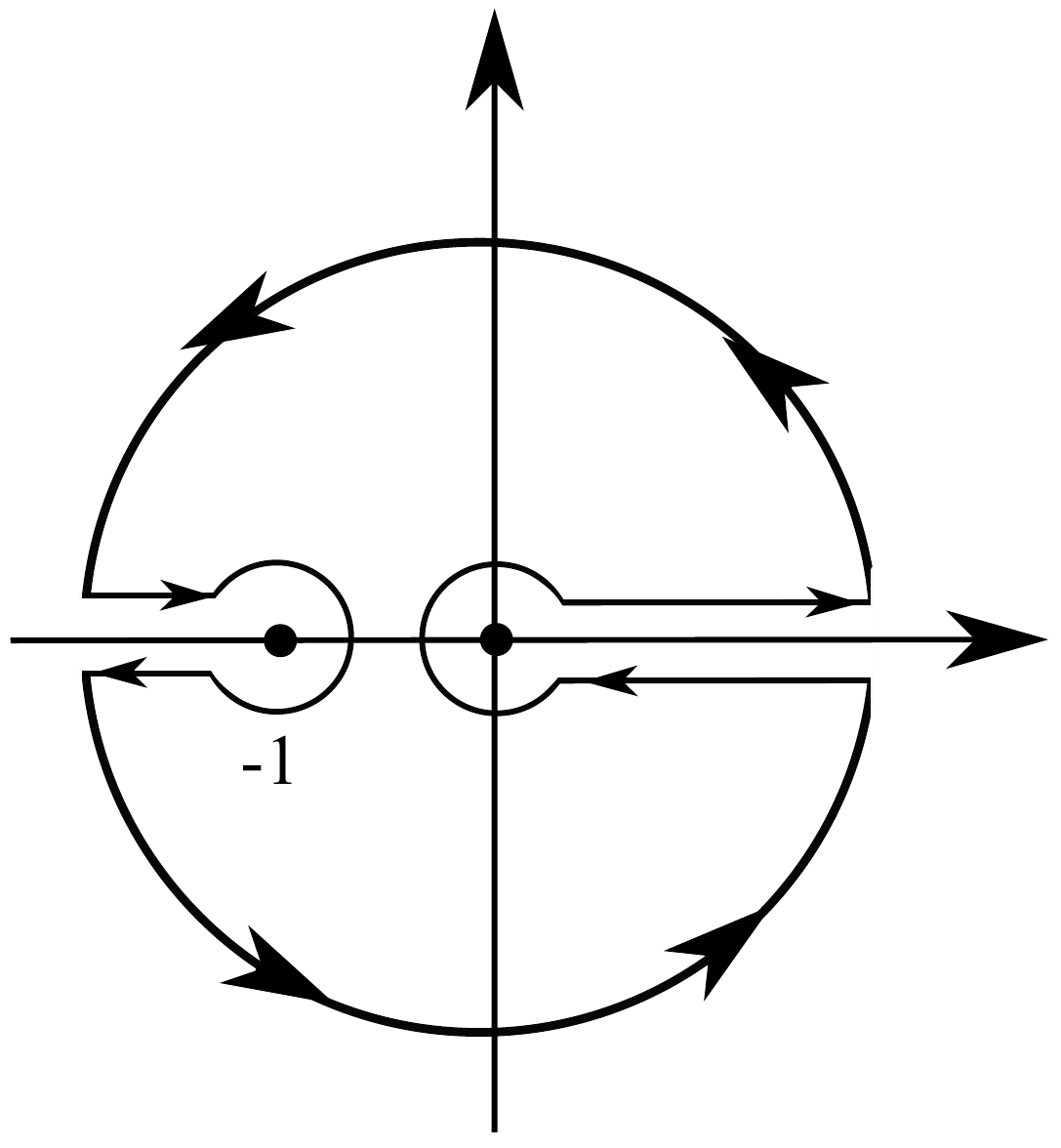}
    \caption{Contour to solve the Integral of the problem suggested in Equation \ref{Eq:054}.}
    \label{Fig:04}
\end{figure}

\begin{figure}[htb!]
    \centering
    \includegraphics[scale=0.45]{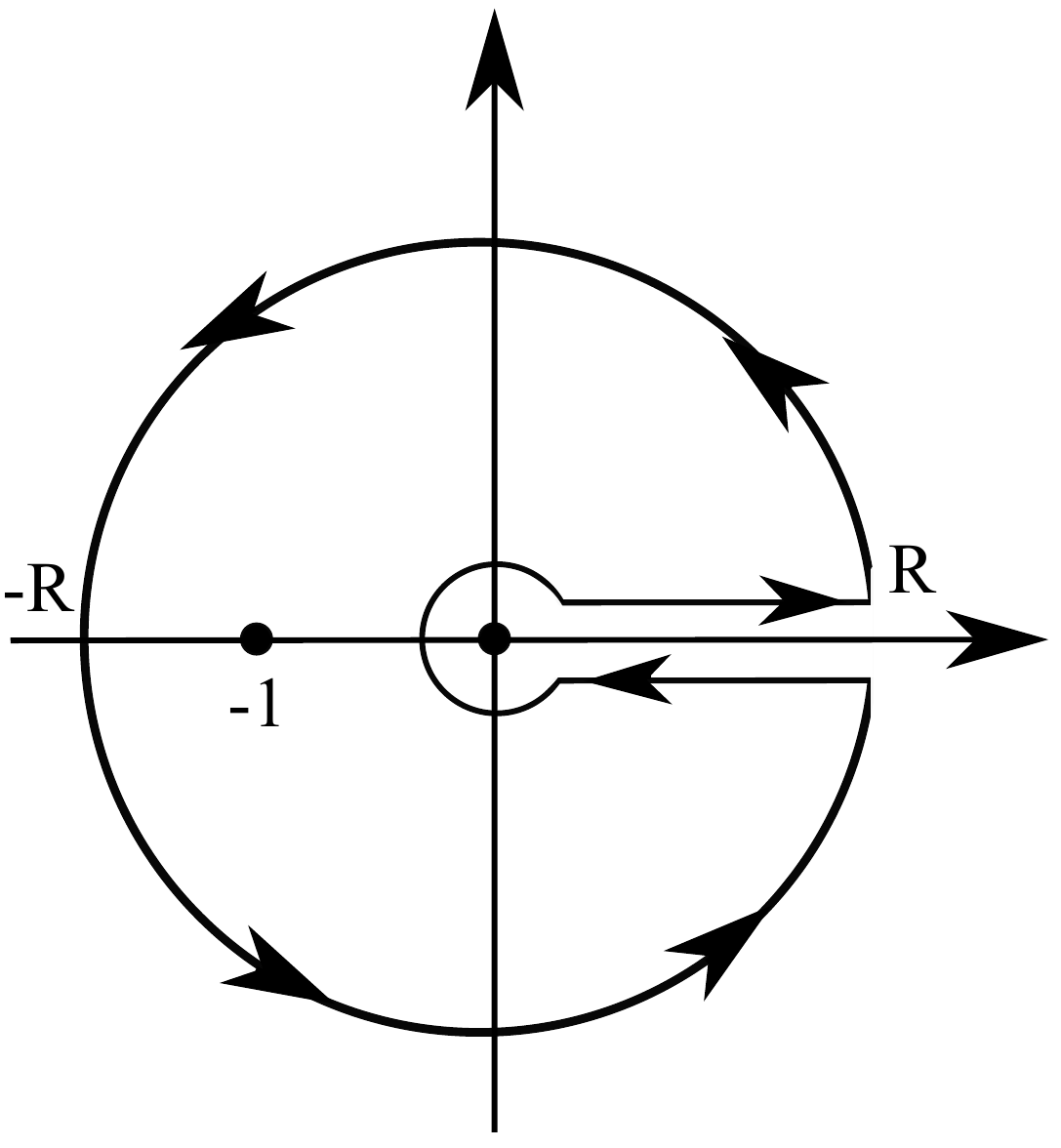}
    \caption{New contour suitable for solving the problem suggested in Equation \ref{Eq:054}.}
    \label{Fig:05}
\end{figure}

\textbf{Resolution:}
Let's consider the following integral:
\begin{eqnarray}
\oint_{C} \frac{e^{a ln(z)}}{(z^{2} + b^{2})^{2}}dz,
\label{Eq:054}
\end{eqnarray}
therefore, considering the appropriate contour (Figure \ref{Fig:04}) for solving the integral \ref{Eq:054}, we therefore have:
\begin{eqnarray}
\int_{-\pi}^{0} \frac{e^{a ln(Re^{i\theta})}iRe^{i\theta}}{(Re^{i\theta} + 1)^{2}}d\theta + \int_{0}^{\pi} \frac{e^{a ln(Re^{i\theta})}iRe^{i\theta}}{(Re^{i\theta} + 1)^{2}}d\theta +  \int_{-(1 - \epsilon)}^{0} \frac{e^{a ln(x e^{i\pi})}e^{i\pi}}{(x^{2 e^{i2\pi}} + 1)}dx +  \int_{R}^{-(1 - \epsilon)} \frac{e^{a ln(x e^{i\pi})}e^{i\pi}}{(x^{2 e^{-i2\pi} + 1)}}dx + \nonumber \\
\int_{\epsilon}^{R} \frac{e^{a ln(x)}dx}{(x^{2} + 1)} + \int_{R}^{\epsilon} \frac{e^{a ln(x)e^{i2\pi}}e^{i2\pi}dx}{(x^{2} + 1)} + \int_{\pi}^{-\pi} \frac{e^{a ln(-1 + \epsilon e^{i\theta})}i \epsilon e^{i\theta}dx}{[(-1 + \epsilon e^{i\theta})^{2} + 1]} + \int_{2\pi}^{0} \frac{e^{a ln(\epsilon e^{i\theta})}i \epsilon e^{i\theta}d\theta}{(\epsilon^{2}e^{i2\theta} + 1)} = 0,
\label{Eq:055}
\end{eqnarray}
considering the limit of Equation \ref{Eq:055} for $ \epsilon \to 0$ and $R \to \infty$, we therefore have:
\begin{eqnarray}
\left| \int_{-\pi}^{0} \frac{e^{a ln(Re^{i\theta})}iRe^{i\theta}d\theta}{(R e^{i\theta} + 1)^{2}} \right| \le \int_{-\pi}^{0} \frac{R^{(a + 1)}d\theta}{(R^{2} - 1)^{2}} ~ \therefore ~ lim_{R \to \infty} \left( \int_{-\pi}^{0} \frac{R^{(a + 1)}d\theta}{(R^{2} - 1)^{2}} \right) = 0 ~ \therefore ~ \nonumber \\
~ \therefore ~ \int_{-\pi}^{0} \frac{e^{a ln(Re^{i\theta})}iRe^{i\theta}d\theta}{(R e^{i\theta} + 1)^{2}} = 0 , ~~ e ~~ ~ \therefore ~ lim_{R \to \infty} \left( \int_{0}^{\pi} \frac{e^{a ln(Re^{i\theta})}iRe^{i\theta}d\theta}{(R e^{i\theta} + 1)^{2}} \right) = 0,
\label{Eq:056}
\end{eqnarray}
so the integral:
\begin{eqnarray}
\int_{\pi}^{-\pi} \frac{e^{a ln(-1 + \epsilon e^{i\theta})}i \epsilon e^{i\theta}d\theta}{(-1 + \epsilon e^{i \theta})^{2} + 1},
\label{Eq:057}
\end{eqnarray}
can be calculated, using the residue theorem, considering the integral:
\begin{eqnarray}
\oint_{C} \frac{e^{a ln(z)}dz}{(z^{2} + 1)}
\label{Eq:058}
\end{eqnarray}
on the contour, where $z = (-1 + \epsilon e^{i\theta})$, orienting clockwise, so:
\begin{eqnarray}
\int_{\pi}^{-\pi} \frac{e^{a ln(-1 + \epsilon e^{i\theta})}i \epsilon e^{i\theta}d\theta}{(-1 + \epsilon e^{i \theta})^{2} + 1} = - 2\pi i Res f(z) ~ \therefore ~ \nonumber \\
~ \therefore ~ f(z) = \frac{e^{a ln(x)}}{(z + 1)}, ~~ where ~~ -1 ~~ is ~~ the ~~ pole ~~ of ~~ order ~~ 2 ~ \therefore ~ Res_{z = -1} = a z^{(a -1)}|_{z=-1} = a e^{(a - 1)ln(-1)} = \nonumber \\
~ \therefore ~ a e^{(a -1) i \pi} = -a e^{ia\pi} ~ \therefore ~ \int_{\pi}^{-\pi} \frac{e^{a ln(-1 + \epsilon e^{i\theta})}i \epsilon e^{i\theta}d\theta}{(-1 + \epsilon e^{i \theta})^{2} + 1} = - 2\pi i (-a e^{ia\pi}) = 2 i \pi a e^{ia\pi} ~ \therefore ~ \nonumber \\
~ \therefore ~ \left| \int_{2\pi}^{0} \frac{e^{a ln(\epsilon e^{i\theta})}i \epsilon e^{i\theta} d\theta}{(\epsilon^{2} e^{i 2 \theta} + 1)} \right| \le \int_{2\pi}^{0} \left| \frac{e^{a(ln(\epsilon) + i\theta)} i \epsilon e^{i\theta}d\theta}{(\epsilon^{2} e^{i 2 \theta} + 1)} \right| = \int_{2\pi}^{0} \frac{|\epsilon^{a}e^{ia\theta} i \epsilon e^{i\theta}| d\theta}{|\epsilon^{2} e^{i 2 \theta} + 1|} \le \int_{2\pi}^{0} \frac{\epsilon^{(a + 1)}}{|\epsilon^{2} - 1|} ~ \therefore ~ \nonumber \\
~ \therefore ~  lim_{\epsilon \to 0} \left( \int_{2\pi}^{0} \frac{|\epsilon^{a} e^{i a \theta} i \epsilon e^{i\theta}| d\theta}{|\epsilon^{2} e^{i2\theta}  + 1|} \right) \le \int_{2\pi}^{0} \frac{\epsilon^{(a + 1)} d\theta}{|\epsilon^{2} - 1|} ~ \therefore ~  lim_{\epsilon \to 0} \left( \frac{\epsilon^{(a + 1)} d\theta}{|\epsilon^{2} - 1|} \right) = lim_{\epsilon \to 0} \left( \frac{\epsilon^{(a + 1)} d\theta}{ 1 - \epsilon^{2}} \right) = 0 ~ \therefore ~ \nonumber \\
~ \therefore ~  0 \le lim_{\epsilon \to 0} \left| \int_{2\pi}^{0} \frac{e^{a ln(\epsilon e^{i\theta})}i \epsilon e^{i\theta} d\theta}{(\epsilon^{2} e^{i 2 \theta} + 1)}  \right| \le lim_{\epsilon \to 0} \left( \int_{2\pi}^{0} \frac{\epsilon^{(a + 1)} d\theta}{(1 - \epsilon^{2})} \right), 
\label{Eq:059}
\end{eqnarray}
So, we can conclude that:
\begin{eqnarray}
lim_{\epsilon \to 0} \left(   \int_{2\pi}^{0} \frac{e^{a ln(\epsilon e^{i\theta})}i \epsilon e^{i\theta} d\theta}{(\epsilon^{2} e^{i 2 \theta} + 1)}\right) = 0 ~ \therefore ~ \int_{0}^{\infty} \frac{x^{a}dx}{(x^{2} + 1)} + \int_{\infty}^{0} \frac{e^{a ln(x e^{i 2 \pi})} e^{i 2 \pi} dx}{(x^{2} + 1)} + 2 i \pi a e^{-i a \pi} = 0 ~ \therefore ~ \nonumber \\ 
\int_{0}^{\infty} \frac{x^{a}dx}{(x^{2} + 1)} + \int_{\infty}^{0} \frac{e^{a(ln(x) + i2\pi)}}{(x^{2} + 1)} dx = - 2 i \pi a e^{-i a \pi} ~ \therefore ~ (1 - e^{(i a 2 \pi)})\int_{0}^{\infty} \frac{x^{a}}{(x^{2} + 1)}dx = - 2 i \pi a e^{i a \pi}
\label{Eq:060}
\end{eqnarray}
\begin{eqnarray}
~ \therefore ~ \boxed{ \int_{0}^{\infty} \frac{x^{a} dx}{(x^{2} + 1)} = -\frac{2 i \pi a e^{(i a \pi)}}{ (1 - e^{(i a 2 \pi)}) } \pi a =  \frac{\pi a}{ sen (\pi a)}}
\label{Eq:061}
\end{eqnarray}
It is important to note that the student can also solve the integral (Equation \ref{Eq:054}), using the outline represented in Figure \ref{Fig:05}, with the same degree of difficulty.

Let's consider the seventh integral (g), Equation \ref{Eq:062}:
\begin{eqnarray}
(g) ~~~\int_{0}^{2 \pi } \frac{d x}{( a \pm b cos(x))} = \int_{0}^{2\pi} \frac{dx}{( a \pm b sen(x))} = \frac{2 \pi}{\sqrt{(a^{2} - b^{2}})}, ~~ where ~~ (a > |b|),
\label{Eq:062}
\end{eqnarray}

\textbf{Resolution:}
let's consider the integral:
\begin{eqnarray}
\oint_{C} \frac{d z}{( a \pm b cos(z))}
\label{Eq:063}
\end{eqnarray}
We have to:
\begin{eqnarray}
y = e^{ i x} = dy = i e^{i x}dx = dx = \frac{dy}{i y} ~ \therefore ~ \oint_{C} \frac{dy}{iy \left[ a \pm b \left(  \frac{y^{2} + 1}{2y}\right)  \right]} =  \oint_{C} \frac{dy}{\frac{i}{2} \left( 2ay \pm by^{2} \pm b\right)} = \nonumber\\
=  \oint_{C} \frac{2dy}{i \left( by^{2} \pm 2ay + b\right)} ~ \therefore ~ \oint_{C} \frac{\pm 2dy}{i \left( by^{2} \pm 2ay + b\right)} = - 2 \pi i Res f(z) ~ \therefore ~  \nonumber \\ 
~ \therefore ~ by^{2} \pm 2ay + b = 0 ~ \therefore ~ \Delta = 4a^{2} + 4b^{2} ~\therefore ~  \sqrt{\Delta} = 2\sqrt{a^{2} + b^{2}} ~\therefore ~ \nonumber \\
~\therefore ~ y = \frac{\pm 2a \pm 2 \sqrt{a^{2} + b^{2}}}{2b} = \pm \frac{a}{b} \frac{\sqrt{a^{2} + b^{2}}}{b} ~\therefore ~   ~~ Polo (1): \frac{a}{b} + \frac{\sqrt{a^{2} + b^{2}}}{b}, ~~ Polo (2): -\frac{a}{b} + \frac{\sqrt{a^{2} + b^{2}}}{b} ~ \therefore ~  \nonumber \\ 
~ \therefore ~ \int_{0}^{2\pi} \frac{2dy}{i(by^{2} + 2ay + b)} = 2 \pi i Res f(z),  considering: ~~ y = \frac{a}{b} + \frac{\sqrt{a^{2} b^{2}}}{b} ~ \therefore ~  \nonumber \\
~ \therefore ~ \int_{0}^{2\pi} \frac{-2dy}{i(by^{2} - 2ay + b)} = 2 \pi i Res f(z),  considering: ~~ y = -\frac{a}{b} + \frac{\sqrt{a^{2} b^{2}}}{b}  ~ \therefore ~  \nonumber \\ 
Res f(z)_{y = -\frac{a}{b} + \frac{\sqrt{a^{2} b^{2}}}{b}} = \frac{2}{i(2by + 2a)} = \frac{2}{i2b \left( -\frac{a}{b} + \frac{\sqrt{a^{2} + b^{2}}}{b} \right) + 2a} = \frac{2}{i 2 \sqrt{a^{2} + b^{2}}} = \frac{1}{i\sqrt{a^{2} + b^{2}}} \therefore ~  \nonumber \\ 
Res f(z)_{y = \frac{a}{b} - \frac{\sqrt{a^{2} b^{2}}}{b}} = \frac{-2}{i(2by - 2a)} = \frac{-2}{-i2\sqrt{a^{2} + b^{2}}} = \frac{1}{i\sqrt{a^{2}+b^{2}}} ~ \therefore ~  \nonumber \\ 
~ \therefore ~ \boxed{ \int_{0}^{\infty} = \frac{dx}{a \pm b cos(x)} = \frac{2 \pi}{\sqrt{a^{2} + b^{2}}}} ~~~~~~~~ (64) ~~~
\label{Eq:064}
\end{eqnarray}
Considering now the integral on the same contour as before, we then have:
\begin{eqnarray}
\int_{0}^{2\pi} = \frac{dx}{(a \pm b sen(x))} ~ \therefore ~ \oint_{C}\frac{dz}{(a \pm sen(z))}
\label{Eq:065}
\end{eqnarray}
\begin{eqnarray}
 \oint_{C}\frac{dz}{(a \pm sen(z))} = \oint_{C} \frac{dy}{iy \left[ a \pm b \left(  \frac{y^{2} - 1}{2 i y} \right) \right]} = \oint_{C} \frac{2dy}{2iay \pm b (y^{2} - 1) } = \oint_{C} \frac{2dy}{\pm by^{2} + 2 i a y \pm b } = ~ \therefore ~  \nonumber \\ 
 ~ \therefore ~ \boxed{ = \oint_{C} \frac{\pm 2dy}{by^{2} \pm 2 i a y - b}}.
 \label{Eq:066}
 \end{eqnarray}
So let's get the roots:
\begin{eqnarray}
 by^{2} \pm 2 i a y - b = 0 ~ \therefore ~ \sqrt{\Delta} = 2i\sqrt{a^{2} + b^{2}} ~ \therefore ~ y' = i \left( -\frac{a}{b} + \frac{\sqrt{a^{2} - b^{2}}}{b} \right) ~~ e ~~ \nonumber \\
  y'' = i \left( \frac{a}{b} - \frac{\sqrt{a^{2} + b^{2}}}{b} \right)
  \label{Eq:067}
\end{eqnarray}
Therefore, the integral (Equation \ref{Eq:066}) becomes:
\begin{eqnarray}
\oint_{C} \frac{2dy}{(by^{2} + 2 i a y - b)} = 2 \pi Res f(z) _{y=\left( -\frac{a}{b} + \frac{\sqrt{a^{2} - b^{2}}}{b}\right)i}
\label{Eq:068}
\end{eqnarray}
and
\begin{eqnarray}
\oint_{C} \frac{-2dy}{(by^{2} - 2 i a y - b)} = 2 \pi Res f(z) _{y=\left( \frac{a}{b} - \frac{\sqrt{a^{2} + b^{2}}}{b}\right)i}
\label{Eq:069}
\end{eqnarray}
therefore the result for the residue of the integral (Equation \ref{Eq:069}), is:
\begin{eqnarray}
 Res f(z) _{y=\left( -\frac{a}{b} + \frac{\sqrt{a^{2} - b^{2}}}{}\right)} = \frac{2}{i2b \left( -\frac{a}{b} + \frac{\sqrt{a^{2} + b^{2}}}{b}\right) + 2 i a} = \frac{1}{i \sqrt{a^{2} + b^{2}}} ~ \therefore ~ \nonumber \\
 ~ \therefore ~ Res f(z) _{y=\left( \frac{a}{b} - \frac{\sqrt{a^{2} - b^{2}}}{b}\right)} = \frac{-2}{i2b \left( \frac{a}{b} - \frac{\sqrt{a^{2} - b^{2}}}{b}\right) - 2 i a} = \frac{1}{i \sqrt{a^{2} + b^{2}}},
 \label{Eq:070}
\end{eqnarray}
so,
\begin{eqnarray}
\int_{0}^{2 \pi} \frac{d\theta}{(a \pm b sen(\theta))} = \frac{2 \pi}{\sqrt{a^{2} - b^{2}}} ~ \therefore ~ \boxed{ \int_{0}^{2 \pi} \frac{d\theta}{(a \pm b sen(\theta))} = \int_{0}^{2 \pi} \frac{d\theta}{(a \pm b cos(\theta))} = \frac{2 \pi}{\sqrt{a^{2} + b^{2}}}}
\label{Eq:071}
\end{eqnarray}

Let's consider the eighth integral (h), Equation \ref{Eq:072}:
\begin{eqnarray}
(h) ~~~\int_{-\infty}^{\infty} \frac{e^{ax}}{(1 + e^{x})}dx =  \frac{\pi}{sen(a \pi )}, ~~ with ~~ |a| < 1 
\label{Eq:072}
\end{eqnarray}

\textbf{Resolution:}
let's consider the contour: $x$ from $-R$ to $R$; $y$ from $0$ to $2 \pi$, so we have the integral:
\begin{eqnarray}
\oint_{C} \frac{e^{a z}}{(1 + e^{z})}dz, 
\label{Eq:073}
\end{eqnarray}
therefore, we have:
\begin{eqnarray}
\int_{-R}^{R} \frac{e^{a x}}{(1 + e^{x})}dx + \int_{-R}^{R} \frac{e^{ax} e^{2\pi a i}dx}{(1 + e^{x}e^{2\pi i})} + \int_{0}^{2\pi} \frac{e^{a(R + iy)} i dy }{(1 + e^{R + iy})} +  \int_{2\pi}^{0} \frac{e^{a(-R + iy)} i dy }{(1 + e^{-R + iy})} = 2 \pi i Res f(z), 
\label{Eq:074}
\end{eqnarray}
considering the limit of the expression above when $R \to \infty $, we get:
\begin{eqnarray}
lim_{R \to \infty} = \left( \int_{-R}^{R} \frac{e^{a x}}{(1 + e^{x})}dx \right) = \left( \int_{-\infty}^{\infty} \frac{e^{a x}}{(1 + e^{x})}dx \right)  ~ \therefore ~ ~~~~~~~~~~\nonumber \\
~ \therefore ~ lim_{R \to \infty} = \int_{-R}^{R} \frac{e^{ax}e^{2\pi a i}}{(1 + e^{x})} = \int_{-\infty}^{\infty} \frac{e^{ax}e^{2\pi a i}}{(1 + e^{x})} ~ \therefore ~ ~~~~~~~~~~  \nonumber \\
\left| \int_{0}^{2\pi} \frac{e^{a(R + iy)} i dy}{(1 + e^{(R + iy)})}  \right| \le \int_{0}^{2\pi}  \left| \frac{e^{a(R + iy)} i dy}{(1 + e^{(R + iy)})} \right|, ~~ therefore, ~~ | 1 + e^{(R + iy)} \ge \left| |e^{(R + iy)}| - |1|  \right| = e^{R} - 1 ~ \therefore ~ ~~~~~~~~~~ \nonumber \\
\int_{0}^{2\pi} \frac{e^{a(R + iy)} i dy}{(1 + e^{(R + iy)})} \le \int_{0}^{2\pi} \frac{e^{aR} dy}{(e^{R} - 1)} ~ \therefore ~ lim_{R \to \infty} \left( \int_{0}^{2\pi} \frac{e^{aR} dy}{(e^{R} - 1)} \right) ~~ (|a| < 1) ~\therefore ~ ~~~~~~~~~~ \nonumber \\
~ \therefore ~ 0 \le lim_{R \to \infty} \left| \int_{0}^{2\pi} \frac{e^{a(R + iy)} i dy}{(1 + e^{(R + iy)})} \right| \le lim_{R \to \infty} \left( \int_{0}^{2\pi} \frac{e^{aR} dy}{(e^{R} - 1)}\right) ~ \therefore ~ lim_{R \to \infty} \left( \int_{0}^{2\pi} \frac{e^{a(R + iy)} i dy}{(1 + e^{(R + iy)})}\right) = 0 ~ \therefore ~  \nonumber \\
~ \therefore ~ \left| \int_{0}^{2\pi} \frac{e^{a(-R + iy)} i dy}{(1 + e^{(-R + iy)})}  \right| \le \int_{0}^{2\pi} \left|  \frac{e^{a(-R + iy)} i dy}{(1 + e^{(-R + iy)})}  \right|  ~ \therefore ~ |1 + e^{-R + iy}| \le |1 - |e^{(-R + iy)}|| = |1 - e^{-R}|  ~ \therefore ~ ~~~~~~~~~~ \nonumber \\
~ \therefore ~ \int_{0}^{2\pi} \left| \frac{e^{a(-R + iy)} i dy}{(1 + e^{(-R + iy)})} \right| \le \int_{0}^{2\pi} \frac{e^{-aR}}{|1 - e^{-R}|} ~ \therefore ~ lim_{R \to \infty} \left(  \int_{2\pi}^{0} \frac{e^{-R} dy}{|1 - e^{-R}|}\right) = lim_{R \to \infty}  \left( \int_{2\pi}^{0} \frac{e^{-aR} dy}{(1 - e^{-R})} \right)= 0 ~ \therefore ~ \nonumber \\
~ \therefore ~ 0 \le lim_{R \to \infty} \left| \int_{2\pi}^{0} \frac{e^{a(-R + iy)} i dy}{(1 + e^{(-R + iy)})}   \right| \le  lim_{R \to \infty} \left( \int_{2\pi}^{0} \frac{e^{-aR}}{(1 - e^{-R})} \right) ~ \therefore ~ ~~~~~~~~~~ \nonumber \\
~ \therefore ~ \boxed{ lim_{R \to \infty} \left( \int_{2\pi}^{0} \frac{e^{a(-R + iy)} i dy}{(1 + e^{-R + iy})}  \right) = 0}, ~~~~~~~~~~
\label{Eq:075}
\end{eqnarray}
therefore,
\begin{eqnarray}
\int_{-\infty}^{\infty} \frac{e^{ax} }{(1 + e^{x}) }dx + \int_{-\infty}^{\infty} \frac{e^{ax}e^{2\pi a i }}{(1 + e^{x})}dx = 2 \pi i Res f(z) ~ \therefore ~ \nonumber \\
~ \therefore ~ Pole: ~~ x = i\pi \nonumber \\
~ \therefore ~ Res f(z) = \frac{P(z)}{Q(z)}, ~~ onde, ~~ P(z) = e^{ax} ~~ e ~~ Q(z) = 1 + e^{x} ~ \therefore ~ Res f(z) = \frac{e^{a i \pi}}{e^{i\pi}} = e^{i a \pi} ~ \therefore ~ \nonumber \\
~ \therefore ~ \int_{-\infty}^{\infty} \frac{e^{ax}}{(1 + e^{x})}dx - cos(2\pi a) \int_{-\infty}^{\infty} \frac{e^{ax}}{(1 + e^{x})}dx - i sen(2 \pi a) \int_{-\infty}^{\infty} \frac{e^{ax}}{(1 + e^{x})}dx = ~ \therefore ~ \nonumber \\
 = 2 \pi i (cos(a \pi) + i sen(a \pi)) ~ \therefore ~ (1 - cos(2\pi a)) \int_{-\infty}^{\infty}\frac{e^{ax}}{(1 + e^{x})}dx - i sen(2 \pi a) \int_{-\infty}^{\infty} \frac{e^{ax}}{(1 + e^{x})}dx ~ \therefore ~ \nonumber \\
~ \therefore ~ sen(2\pi a) \int_{-\infty}^{\infty}  \frac{e^{ax}}{(1 + e^{x})}dx = 2 \pi cos(\pi a) ~ \therefore ~ \int_{-\infty}^{\infty} \frac{e^{ax}}{(1 + e^{x})}dx = \frac{2 \pi cos(\pi a)}{2 cos(\pi a) sen(\pi a)} ~ \therefore ~ \nonumber \\
~ \therefore ~ \boxed{ \int_{-\infty}^{\infty} \frac{e^{ax}}{(1 + e^{x})}dx = \frac{\pi}{sen(\pi a)}}
\label{Eq:076}
\end{eqnarray}

Let's consider the ninth integral (i), Equation \ref{Eq:077}:
\begin{eqnarray}
(i) ~~~\int_{0}^{2\pi} \frac{cos^{2}(3x)}{(5 - 4cos(2x))}dx =  \frac{3\pi}{8}
\label{Eq:077}
\end{eqnarray}

\textbf{Resolution:}
let's consider the integral on the circular contour of unit radius, then:
\begin{eqnarray}
\oint_{C}  \frac{cos^{2}(3z)}{(5 - 4cos(2z))}dz, 
\label{Eq:078}
\end{eqnarray}
therefore, we have:
\begin{eqnarray}
\oint_{C}  \frac{cos^{2}(3z)}{(5 - 4cos(2z))}dz = \oint_{C}  \frac{\frac{1}{2}(1 + cos(6x))}{(5 - 4cos(2x))}dx = ~ \therefore ~ 2x = u \to du = 2dx \to dx = \frac{du}{2} ~ \therefore ~ \nonumber \\
~ \therefore ~ = \frac{1}{4} \oint_{C} \frac{1 + cos(3u)}{5 - 4 cos(u)}du ~ \therefore ~ on ~~ the ~~ contour: ~~ \left( \frac{1}{4} \int_{0}^{2\pi} \frac{1 + cos(3u)}{5 - 4 cos(u)}du \right) ~ \therefore ~ \nonumber \\
~ \therefore ~ cos(3u) = \frac{y^{3} + y^{-3}}{2} = \frac{y^{6} + 1}{2y^{3}} ~ \therefore ~ cos(u) = \frac{y^{2} + 1}{2y} ~ \therefore ~ du = \frac{dy}{idy} ~ \therefore ~ \frac{1}{4}\left( \frac{1 + \frac{(y^{6} + 1)}{2y^{3}}}{5 - 4 \frac{y^{2} + 1}{2y}} \right) = ~ \therefore ~ \nonumber \\
~ \therefore ~ = \frac{1}{4}\left(   \frac{y^{6} + 2y^{3}+1}{y^{2}(5y - 2y^{2} -2)}\right) ~ \therefore ~ \frac{i}{4}\int_{0}^{2\pi} \frac{1 + cos(3u)}{(5 - 4cos(u))}du = \frac{i}{8}\int_{0}^{2\pi}\frac{y^{6} + 2y^{3} + 1}{y^{3}(y^{2} -\frac{5}{2}y + 1)}dy ~ \therefore ~ \nonumber \\
~ \therefore ~ as ~~ raízes ~~ são: ~~ y^{2} -\frac{5}{2}y + 1 = 0 ~ \therefore ~ \Delta = \frac{25}{4} - = \frac{25 - 16}{4} = \frac{9}{4} ~ \therefore ~ \sqrt{\Delta} = \frac{3}{2}, ~ \therefore ~ \nonumber \\
~ \therefore ~ y' = \frac{\frac{5}{2} - \frac{3}{2}}{2} = \frac{1}{2} ~ \therefore ~ y'' = \frac{\frac{5}{2} + \frac{3}{2}}{2} = 2
\label{Eq:079}
\end{eqnarray}
So, let's consider the internal poles of the contour: $0$ (where $3$) and $\frac{1}{2}$, so we have:
\begin{eqnarray}
f(z) = \frac{i}{8} \frac{y^{6} + 2y^{3} +1}{(y^{2} - \frac{5}{2} + 1)} ~ \therefore ~ f'(z) = \frac{i}{8}\frac{(y^{2} - \frac{5}{2}y + 1)(6y^{5} + 6y^{2}) - (2y - \frac{5}{2})(y^{6} + 2y^{3} + 1)}{(y^{2} - \frac{5}{2}y + 1)^{2}} ~ \therefore ~ \nonumber \\
f''(z)=\frac{i}{8}\frac{\left[ (2y -\frac{5}{2})(6y^{5}+6y^{2}) + (y^{2} - \frac{5}{2}y + 1)(30y^{4} + 12y) - 2(y^{6} + 2y^{3} + 1) - (2y - \frac{5}{2})(6y^{5} + 6y^{2}) \right]}{(y^{2} - \frac{5}{2} + 1)^{2}} ~ \therefore ~ \nonumber \\
 ~ \therefore ~ f''(z) = \frac{i}{8}\frac{2(2y - \frac{5}{2})}{(y^{2} - \frac{5}{2}y + 1)^{3}} \left[ (y^{2} - \frac{5}{2}y +1)(6y^{5} + 6y^{2}) - (2y - \frac{5}{2})(y^{6}+2y^{3}+1) \right]  ~ \therefore ~ \nonumber \\
 Res_{z=0} = \frac{f''(0)}{2!} = \frac{i}{16}(1)^{-3} \left[(1)(0) - (2)(1)\right] - \frac{i}{8}(-5) \left[  (1)(0) + \left(\frac{5}{2}\right)(1)\right] ~ \therefore ~ \nonumber \\
 ~ \therefore ~ \boxed{ Res_{z=0} = -\frac{i}{16} + \frac{i}{16} \frac{25}{2} = -\frac{i}{16} + \frac{25 i}{32} = \frac{21 i}{32}} ~ \therefore ~ \nonumber \\
 Res_{z=\frac{1}{2}} = \frac{i}{8} \frac{(y^{6} + 2y^{3} + 1)}{(y^{3}(y -2))} = \frac{i}{8}\frac{\frac{1}{64} + 2\frac{1}{8} + 1}{\frac{1}{8}(-\frac{3}{2})} = \frac{i}{8} \frac{\frac{1}{64} + \frac{16}{64} + \frac{64}{64}}{-\frac{3}{16}} = \frac{i}{8} \frac{81}{64} \frac{16}{-3} = \frac{-27 i}{32} ~ \therefore ~ \nonumber \\
~ \therefore ~ \boxed{\sum Res f(z) = \frac{21 i}{32} - \frac{27 i}{32} = -\frac{6 i}{32} = -\frac{3 i}{16}} ~ \therefore ~ \nonumber \\
\oint_{C} \frac{i}{8} \frac{(y^{6} + 2y^{3} +1)}{y^{3}(y^{2} - \frac{5}{2}y + 1)} = 2 \pi i \left( \frac{-3i}{16} \right) ~ \therefore ~ \boxed{ \int_{0}^{2\pi} \frac{cos^{2}(3x)}{(5 - 4 cos(2x))}dx = \frac{3 \pi}{8}} ~~~~~~~~~~~~~~ (80)
\label{Eq:080}
\end{eqnarray}

Let's consider the tenth integral (j), Equation \ref{Eq:081}:
\begin{eqnarray}
(j) ~~~\int_{0}^{\infty} e^{-x^{2}} cos(2 b x) dx = \frac{1}{2}\sqrt{\pi}e^{-b^{2}}
\label{Eq:081}
\end{eqnarray}

\textbf{Resolution:}
let's consider the integral on the contour: $x$ from $-a$ to $a$ and $y$ from $0$ to $b$, so we have:
\begin{eqnarray}
\oint_{C}  e^{-z^{2}}e^{2 b i z}dz 
\label{Eq:082}
\end{eqnarray}
in the following contour (Figure \ref{Fig:06}):\begin{figure}[htb!]
    \centering
    \includegraphics[scale=0.45]{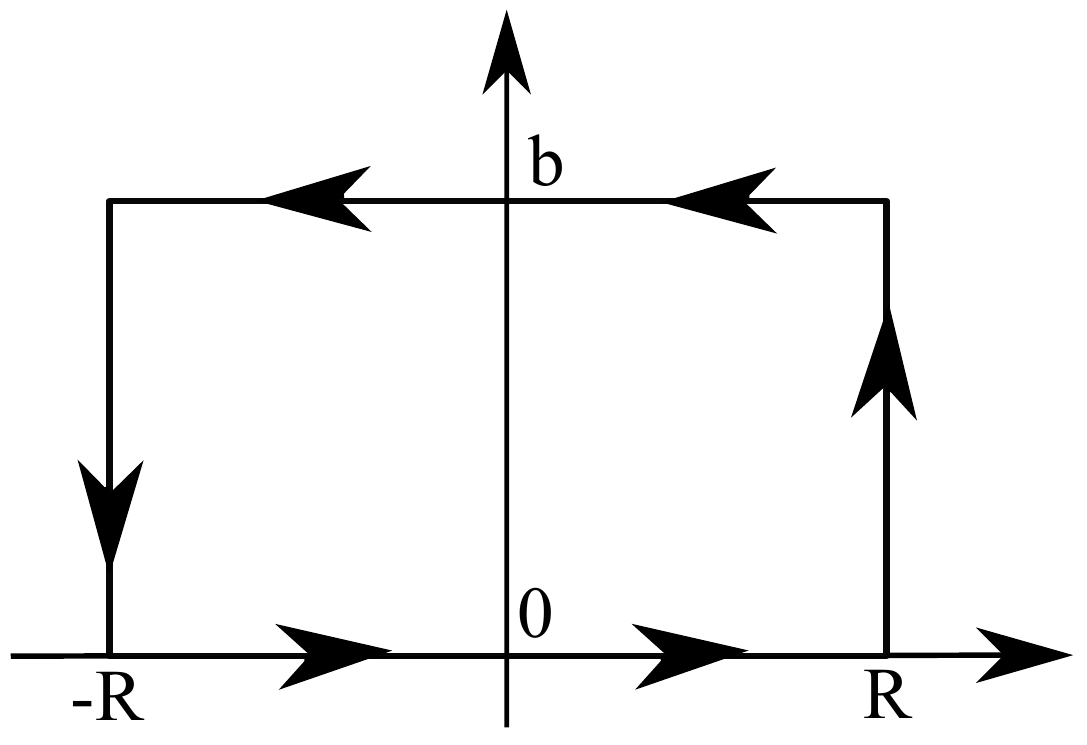}
    \caption{Adequate contour for solving the problem suggested in Equation \ref{Eq:082}.}
    \label{Fig:06}
\end{figure}
therefore, we have to:
\begin{eqnarray}
\int_{-R}^{R} e^{-x^{2}} e^{2 b i x} dx + \int_{-R}^{R} e^{-(x + bi)^{2}} e^{2 b i (x + b i)}dx + \int_{0}^{b} e^{-(-R + iy)^{2}} e^{2 b i (R + iy)} i dy + \nonumber \\
+ \int_{b}^{0} e^{-(-R + iy)^{2}} e^{2 b i (-R + iy)} i dy = 0,
\label{Eq:083}
\end{eqnarray}
considering the limit of Equation \ref{Eq:083}, when $R \to \infty$, we therefore have:
\begin{eqnarray}
lim_{R \to \infty} = \left(  \int_{-R}^{R} e^{- x^{2}} e^{2 b i x} dx \right) = \int_{-\infty}^{\infty} e^{- x^{2}} e^{2 b i x} dx ~ \therefore ~ \nonumber \\
~ \therefore ~ lim_{R \to \infty} = \int_{-R}^{R} e^{-(x + b i)^{2}} e^{2 b i ( x + bi)}dx = \int_{-\infty}^{\infty} e^{(- x^{2} + b^{2} -2bix)} e^{(2bix - 2b^{2})}dx = - e^{-b^{2}}\sqrt{\pi},
\label{Eq:084}
\end{eqnarray}
So let's define:
\begin{eqnarray}
I = \int_{-\infty}^{\infty} e^{-x^{2}}dx ~~~ e ~~~ I = \int_{-\infty}^{\infty} e^{-y^{2}}dy, ~~~ logo: ~~~ \nonumber \\
~ \therefore ~  I.I = \int_{-\infty}^{\infty} \int_{-\infty}^{\infty}  e^{-x^{2}} e^{-y^{2}}dx dy ~ \therefore ~ I^{2} = \int_{-\infty}^{\infty} \int_{-\infty}^{\infty} e^{-(x^{2} + y^{2})}dxdy
\label{Eq:085}
\end{eqnarray}
solving Equation \ref{Eq:085}, we have:
\begin{eqnarray}
x^{2} + y^{2} = \rho^{2} ~~~ e ~~~ dxdy = \rho (d\rho) (d\theta), ~~~ logo:
\label{Eq:086}
\end{eqnarray}
\begin{eqnarray}
I^{2} = \int_{0}^{2\pi} \int_{-\infty}^{\infty} e^{-(\rho^{2})} \rho (d\rho) (d\theta) = \int_{0}^{2\pi} \frac{1}{2} d\theta = \frac{1}{2}(2 \pi) = \sqrt{\pi},
\label{Eq:087}
\end{eqnarray}
so,
\begin{eqnarray}
\left| \int_{0}^{b} e^{-(R + iy)^{2}} e^{2 b i (R + i y)} i dy  \right| \le \int_{0}^{b} \left|  e^{-(R + iy)^{2}} e^{2 b i (R + i y)} i dy  \right| = \int_{0}^{b} e^{-R^{2}e^{(y^{2} - 2 b y)}}dy = 0  ~ \therefore ~ \nonumber \\
~ \therefore ~ lim_{R \to \infty} \left( \int_{0}^{b} e^{-R^{2}e^{(y^{2} - 2 b y)}}dy  \right) = 0 ~ \therefore ~ \nonumber \\
~ \therefore ~ 0 \le \left| \int_{0}^{b} e^{-(R + iy)^{2}} e^{2 b i (R + iy)} i dy   \right| \le lim_{R \to \infty} \left( \int_{0}^{b} e^{-R^{2}} e^{(-y^{2} - 2 b y)}dy  \right) ~ \therefore ~ \nonumber \\
~ \therefore ~  \int_{0}^{b} e^{-(R + iy)^{2}} e^{2 b i (R + i y)} i dy = 0  ~ \therefore ~ \nonumber \\
~ \therefore ~  \left| \int_{b}^{0} e^{-(-R + iy)^{2}} e^{2 b i (-R + i y)} i dy  \right| \le \int_{b}^{0} \left| e^{-(-R + iy)^{2}} e^{2 b i (-R + i y)} i dy  \right| = \int_{b}^{0} \left| e^{(-R^{2} + 2 i R y + y^{2})} e^{(-2 b i R - 2 b y)} i dy\right| = ~ \therefore ~ \nonumber \\
 = \int_{b}^{0} e^{-R^{2}} e^{(y^{2} - 2 b y)} dy ~ \therefore ~ lim_{R \to \infty}  \int_{b}^{0} \left( e^{(y^{2} - 2 b y)} dy   \right) = 0 ~~~~~~~~~
 \label{Eq:088}
 \end{eqnarray}
 \begin{eqnarray}
 lim_{R \to \infty} \left( \int_{b}^{0} e^{-(-R + iy)^{2}} e^{2 b i (-R + i y)} i dy \right) = 0 ~ \therefore ~ \nonumber \\ 
~ \therefore ~ \int_{-\infty}^{\infty} e^{-x^{2}} e^{2 b i x} dx - e^{-b^{2}}\sqrt{\pi} = 0 ~ \therefore ~ \int_{-\infty}^{\infty} e^{-x^{2}} e^{2 b i x} dx = e^{-b^{2}}\sqrt{\pi}
\label{Eq:089}
\end{eqnarray}
logo,
\begin{eqnarray}
\boxed{\int_{0}^{\infty} e^{-x^{2}} cos(2 b x) dx = \int_{0}^{\infty} e^{-x^{2}} e^{2 b i x} dx = \frac{\sqrt{\pi}}{2} e^{-b^{2}}}
\label{Eq:090}
\end{eqnarray}

Let's consider the eleventh integral (l), Equation \ref{Eq:091}:\begin{eqnarray}
(l) ~~~\int_{0}^{\infty} \frac{sen(a x)}{senh(x)} dx
\label{Eq:091}
\end{eqnarray}

\textbf{Resolution:}\\
As a suggestion for solving the integral represented in Equation \ref{Eq:092}, we have:
\begin{eqnarray}
\oint_{C} \frac{e^{(iaz)}}{senh(z)} dz
\label{Eq:092}
\end{eqnarray}
where we will consider for the integral the contour represented in Figure \ref{Fig:07} below.
\begin{figure}[htb!]
    \centering
  \includegraphics[scale=0.55]{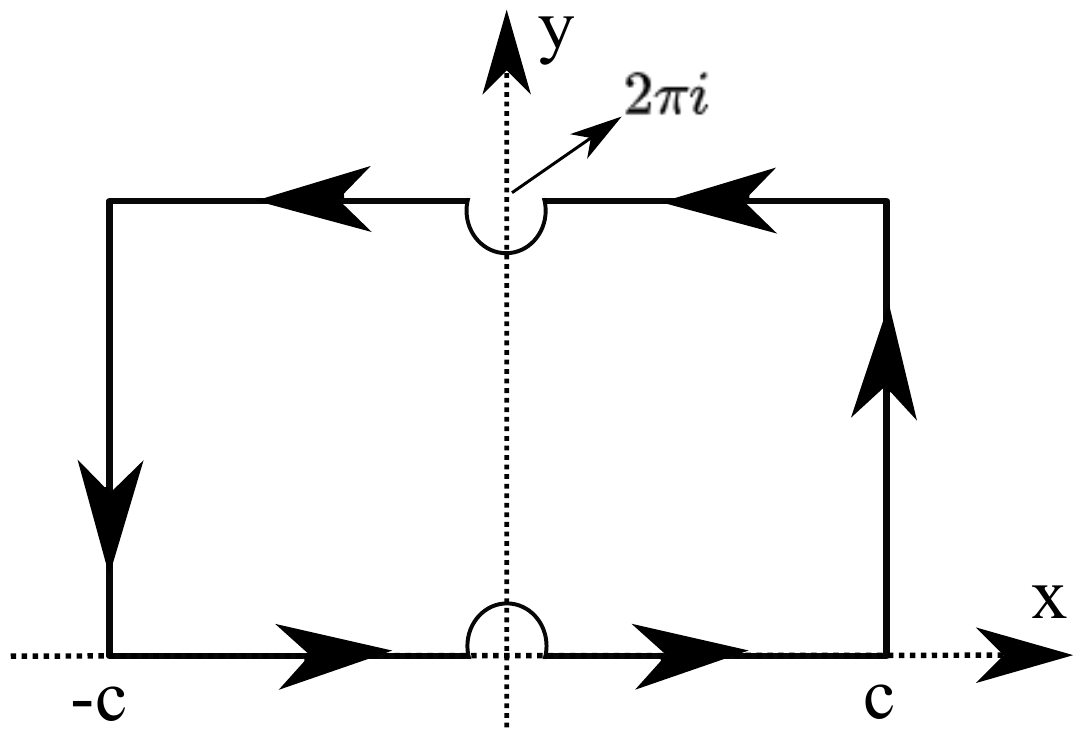}
    \caption{Adequate contour for solving the problem suggested in Equation \ref{Eq:092}.}
   \label{Fig:07}
\end{figure}
\begin{eqnarray}
\int_{0}^{2\pi} \frac{2 e^{i a(c + iy)}}{e^{(c + iy)} - e^{(-c - iy)}}idy + \int_{2\pi}^{0} \frac{2 e^{i a(-c + iy)}}{e^{(-c + iy)} - e^{(c - iy)}}idy + \int_{\pi}^{0} \frac{2 e^{i a \epsilon e^{i \theta}} i \epsilon e^{i\theta}}{e^{\epsilon e^{i\theta}} - e^{-\epsilon e^{i\theta}}}d\theta +  \nonumber \\
 + \int_{0}^{-\pi} \frac{2 e^{i a \epsilon e^{i\theta}} e^{i a 2 \pi i} i \epsilon e^{i\theta}}{e^{2\pi i}(e^{\epsilon e^{i\theta}} - e^{-\epsilon e^{i\theta}})} d\theta + \int_{\epsilon}^{R} \frac{e^{i a x}}{senh(x)}dx + \int_{R}^{\epsilon} \frac{e^{i a x} e^{i a 2 \pi i}}{senh(x)}dx + \nonumber \\
 + \int_{-\epsilon}^{-R} \frac{e^{i a x} e^{i a 2 \pi i}}{senh(x)}dx + \int_{-R}^{-\epsilon} \frac{e^{i a x}}{senh(x)}dx = 2 \pi i Res f(z)
 \label{Eq:093}
\end{eqnarray}
We have to: $e^{+z} - e^{-z} = 0 \Rightarrow e^{z} = e^{i 2\pi}e^{-z} \Rightarrow z =  i2 \pi - z \Rightarrow z = i\pi$, thus:
\begin{eqnarray}
Res_{z = i\pi} f(z) = lim_{z \to i\pi} \left[ (z - i\pi) \frac{e^{i a z}}{senh(z)}  \right] = \frac{e^{i a z}}{cosh(z)}|_{z=i\pi} = \frac{e^{-a\pi}}{cosh(i\pi)} = - e^{-a\pi},
\label{Eq:094}
\end{eqnarray}
calculating the limit of the expression above when $c \to \infty$ and $\epsilon \to 0$, we therefore have:
\begin{eqnarray}
\left| \int_{0}^{2\pi} \frac{2 i e^{i a c} e^{-y}}{(e^{iy} e^{c} - e^{-iy} e^{-c})}dy\right| \le \int_{0}^{2\pi} \left| \frac{2 i e^{i a c} e^{-y}}{(e^{iy} e^{c} - e^{-iy} e^{-c})}dy\right| \le \int_{0}^{2\pi} \frac{e^{-ay}}{|e^{c} - e^{-c}|}dy ~ \therefore ~ \nonumber \\
~ \therefore ~ lim_{c \to \infty} \left( \int_{0}^{2\pi} \frac{e^{-ay}}{|e^{c} - e^{-c}|}dy \right) = lim_{c \to \infty} \left( \int_{0}^{2\pi} \frac{e^{-ay}}{e^{c} - e^{-c}}dy \right) = 0,
\label{Eq:095}
\end{eqnarray}
as:
\begin{eqnarray}
0 \le lim_{c \to \infty} \left|  \int_{0}^{2\pi} \frac{2 e^{i a(c + iy)}}{e^{(c + iy)} - e^{(-c - iy)}}idy \right| \le lim_{c \to \infty} \left(\int_{0}^{2\pi} \frac{e^{-a y}}{e^{c} - e^{-c}}dy  \right), ~~~ deduzimos ~~~ que ~~~ \nonumber \\
lim_{c \to \infty} \left(\int_{0}^{2\pi} \frac{2 e^{i a(c + iy)}i}{[e^{(c + iy)} - e^{(-c -iy)}]}dy  \right) = 0,
\label{Eq:096}
\end{eqnarray}
so,
\begin{eqnarray}
\left|  \int_{2\pi}^{0} \frac{2 e^{ia(-c + iy)}i}{[e^{(-c + iy)} - e^{(c - iy)}]}dy \right| \le \int_{2\pi}^{0} \left|   \frac{2 e^{ia(-c + iy)}i}{[e^{(-c + iy)} - e^{(c - iy)}]}dy \right| \le \int_{2\pi}^{0} \frac{|2e^{-iac} e^{-ay}|}{|e^{-c} - e^{c}|}idy ~ \therefore ~ \nonumber \\
~ \therefore ~ lim_{c \to \infty} \left( \frac{e^{-ay}}{|e^{-c} - e^{c}|}dy \right) = lim_{c \to \infty} \left( \frac{e^{-ay}}{e^{c} - e^{-c}}dy \right) ~ \therefore ~ \boxed{ lim_{c \to \infty} \left(  \frac{2 e^{ia(-c + iy)}i}{[e^{(-c + iy)} - e^{(c - iy)}]}dy\right) = 0} \nonumber \\
~ \therefore ~ lim_{(R \to \infty ~~ and ~~ \epsilon \to 0)} \left( \int_{\epsilon}^{R} \frac{e^{i a x}}{senh(x)} dx + \int_{-R}^{-\epsilon} \frac{e^{i a x}}{senh(x)} dx\right) = \int_{-\infty}^{\infty} \frac{e^{i a x}}{senh(x)}dx ~ \therefore ~ \nonumber \\
~ \therefore ~  lim_{(R \to \infty ~~ and ~~ \epsilon \to 0)} \left( \int_{R}^{\epsilon} \frac{e^{i a x} e^{i a 2 \pi i}}{senh(x)}dx + \int_{-\epsilon}^{-R} \frac{e^{iax} e^{ia2\pi i}}{senh(x)} \right) = (- e^{-a2\pi}) \int_{-\infty}^{\infty} \frac{e^{iax}}{senh(x)}dx ~ \therefore ~ \nonumber \\
~ \therefore ~ lim_{\epsilon \to 0} \left( \int_{0}^{2\pi} \frac{2 e^{i a \epsilon e^{i\theta}} i \epsilon e^{i\theta}}{e^{\epsilon e^{i\theta}} - e^{-\epsilon e^{i\theta}}}d\theta \right) = -\pi i Res_{z=0} f(z) ~ \therefore ~ \nonumber \\ 
~ \therefore ~ Res_{z=0} f(z) = lim_{z \to 0} \left( \frac{z e^{i a z}}{senh(z)} dz\right) = \frac{e^{i a z}}{cosh(z)}|_{z=0} = 1 ~ \therefore ~ \nonumber \\ 
~ \therefore ~ lim_{\epsilon \to 0} \left( \int_{0}^{2\pi} \frac{2 e^{i a \epsilon e^{i\theta}} i \epsilon e^{i \theta}}{e^{\epsilon e^{i\theta}} - e^{-\epsilon e^{i\theta}}}d\theta \right) = \pi i ~ \therefore ~ lim_{\epsilon \to 0} \left( \int_{0}^{\pi} \frac{2 e^{i a \epsilon e^{i \theta}} e^{i a 2\pi i} i \epsilon e^{i\theta}}{e^{2\pi i (e^{\epsilon e^{i\theta}} - e^{- \epsilon e^{i\theta}})}}d\theta \right) = - \pi i Res f(z) ~ \therefore ~ \nonumber \\ 
~ \therefore ~ Res_{z = i 2\pi} f(z) = lim_{z \to i 2 \pi} \left[(z - i2\pi) \frac{e^{i a z}}{senh(z)}  \right] = \frac{e^{i a z}}{cosh(z)}|_{z=i 2\pi} = e^{i a i 2 \pi} = e^{-a 2 \pi} ~ \therefore ~ \nonumber \\ 
 ~ \therefore ~ \boxed{ lim_{\epsilon \to 0} \left( \int_{0}^{\pi} \frac{2 e^{i a \epsilon e^{i \theta}} e^{i a 2\pi i} i \epsilon e^{i\theta}}{e^{2\pi i (e^{\epsilon e^{i\theta}} - e^{- \epsilon e^{i\theta}})}}d\theta \right) = - \pi i e^{-a 2 \pi}} ~ \therefore ~ \nonumber \\ 
 ~ \therefore ~ \int_{-\infty}^{\infty} \frac{e^{e^{i a x}}}{senh(x)}dx - \int_{-\infty}^{\infty} e^{-a 2 \pi} \frac{e^{i a x}}{senh(x)}dx - \pi i (1 + e^{-a 2\pi}) = -2 \pi i e^{-a \pi} ~ \therefore ~ \nonumber \\ 
 ~ \therefore ~ \boxed{ \int_{-\infty}^{\infty} \frac{e^{e^{i a x}}}{senh(x)}dx = \frac{\pi i (e^{-a 2 \pi} - 2 e^{-a\pi} + 1)}{(1 - e^{-a 2 \pi})} = \frac{\pi i (1 - e^{- a \pi})^{2}}{(1 - e^{-a \pi})(1 + e^{- a \pi})} = \frac{i \pi (1 - e^{-a\pi})}{(1 + e^{- a \pi})}}
 \label{Eq:097}
\end{eqnarray}

Therefore, we present the resolution in detail of eleven improper integrals in the Complex Plane, (in the set of Complex numbers $\mathbb{C}$). The integrals presented here in this work, the student who attends the discipline of Complex Variables in the area of Sciences and their Technologies can find the solutions, but without the mathematical development in the process of solving the integrals. Thus, we hope that this work can serve as a theoretical basis in the development of calculations for solving other integrals that were not presented here in this work. Therefore, we present that the resolution of an unsolvable integral on the set of real numbers $\mathbb{R}$, we calculate an integral on a closed contour $C$, when for example $z = \zeta$ is $(i)$ outside the contour $C$ and $(ii)$ inside the contour $C$. So when $z = \zeta$ is outside $C$, then the integral is analytic at all interior points and along the contour $C$, so we apply the Cauchy-Goursat Theorem \cite{cauchy1884oeuvres,stewart1960numerical,avila2008variaveis}. If $z = \zeta$ is inside the contour $C$, then we consider a circle $\Xi$ of radius $\alpha$, where $\Xi$ is totally inside $C$, where we directly apply the Theorem of ` `Integration boundary deformation'', and which involve $i$ holes that may contain isolated or non-isolated singularities. For the student to seek more theoretical information on the subject, he can search in the bibliographical references \cite{cauchy1884oeuvres,stewart1960numerical,avila2008variaveis,fisher1999complex,ash2014complex,narasimhan1971several,berenstein2012complex}. 
\section{Conclusions}

The solution of improper integrals involving problems found in nature that cannot be solved in the set of real numbers ($\mathbb{R}$), can be solved by obtaining complex solutions, which must be treated as a function or complex number ($\mathbb {C}$) of a measured physical quantity. Therefore, we can observe in the process of solving the improper integrals presented here in this work, that we must associate the real and imaginary part or another real quantity derived from the modulus of a complex number with real physical parameters. It is important for students to reflect on solving physical problems found in nature. Such reflection can be understood for example when the real index of refraction of an electromagnetic wave that propagates in a continuous medium, becomes a complex quantity when the absorption of the energy carried by the wave is included. We may also reflect that the actual energy associated with an atomic (or nuclear) energy level becomes complex when the finite lifetime of the energy level is considered. We can also find a complex number when a particle is subjected to the infinite quantum potential well. Therefore, we hope that this work will contribute to the training of students who study Complex Variables in the area of Science and Technologies in the process of solving improper integrals in the complex plane.

\section{Acknowledgements}

This work was supported in part by the Brazilian Agencies CAPES, CNPq, FAPESP and FAPEPI. J.M.S. and N.M.P. acknowledges University of Trento, Italy and Queen Mary University of London, UK.

\bibliographystyle{elsarticle-num}
\bibliography{bibliografia.bib}
\end{document}